\documentclass[aps,prb,twocolumn,amsmath,amssymb,showpacs,superscriptaddress]{revtex4-2}
\usepackage{graphicx}						
\usepackage{epstopdf}
\usepackage{amsmath,amssymb,amsfonts}
\usepackage{textcomp}
\usepackage{hyperref}
\usepackage{gensymb}
\usepackage{verbatim}
\usepackage{amsmath}
\hypersetup{breaklinks=true,colorlinks=true,urlcolor=black}
\usepackage{color}
\usepackage{soul}
\usepackage{float}
\usepackage{gensymb}
\usepackage{caption}
\usepackage{subcaption}
\usepackage[english]{babel}
\usepackage[T1]{fontenc}
\usepackage[utf8]{inputenc}
\usepackage{booktabs}
\DeclareGraphicsExtensions{.jpg,.pdf,.png,.eps}

\begin{document}
\title{Superconductivity and magnetic and transport properties of single-crystalline CaK(Fe$_{1-x}$Cr$_{x}$)$_{4}$As$_{4}$}
\author{M. Xu}
\author{J. Schmidt}
\affiliation{Ames National Laboratory, Iowa State University, Ames, Iowa 50011, USA}
\affiliation{Department of Physics and Astronomy, Iowa State University, Ames, Iowa 50011, USA}

\author{M.~A.~Tanatar}
\affiliation{Ames National Laboratory, Iowa State University, Ames, Iowa 50011, USA}
\affiliation{Department of Physics and Astronomy, Iowa State University, Ames, Iowa 50011, USA}

\author{R.~Prozorov}
\affiliation{Ames National Laboratory, Iowa State University, Ames, Iowa 50011, USA}
\affiliation{Department of Physics and Astronomy, Iowa State University, Ames, Iowa 50011, USA}

\author{S. L. Bud'ko}
\affiliation{Ames National Laboratory, Iowa State University, Ames, Iowa 50011, USA}
\affiliation{Department of Physics and Astronomy, Iowa State University, Ames, Iowa 50011, USA}

\author{P. C. Canfield}
\affiliation{Ames National Laboratory, Iowa State University, Ames, Iowa 50011, USA}
\affiliation{Department of Physics and Astronomy, Iowa State University, Ames, Iowa 50011, USA}
\email[]{canfield@ameslab.gov}

\date{\today}

\begin{abstract}
	Members of the CaK(Fe$_{1-x}$Cr$_{x}$)$_{4}$As$_{4}$ series have been synthesized by high-temperature solution growth in single crystalline form and characterized by X-ray diffraction, elemental analysis, magnetic and transport measurements. The effects of Cr substitution on the superconducting and magnetic ground states of CaKFe$_4$As$_4$ ($T_c$ = 35 K) have been studied. These measurements show that the superconducting transition temperature decreases monotonically and is finally suppressed below 1.8 K as $x$ is increased from 0 to 0.038. For $x$-values greater than 0.012, signatures of a magnetic transition can be detected in magnetic measurements with the associated features in the transport measurements becoming detectable for $x$ $\geq$ 0.038. The magnetic transition temperature increases in a roughly linear manner as Cr substitution increases. A temperature-composition (\textit{T}-\textit{x}) phase diagram is constructed, revealing a half-dome of superconductivity with the magnetic transition temperature, $T^*$, appearing near 22~K for $x$ $\sim$ 0.017 and rising slowly up to 60~K for $x$ $\sim$ 0.077. The $T$-$x$ phase diagrams for CaK(Fe$_{1-x}$$T$$_{x}$)$_4$As$_4$ for $T$ = Cr and Mn are essentially the same despite the nominally different band filling; this is in marked contrast to $T$ = Co and Ni series for which the $T$-$x$ diagrams scale by a factor of two, consistent with the different changes in band filling Co and Ni would produce when replacing Fe. Superconductivity of CaK(Fe$_{1-x}$Cr$_{x}$)$_{4}$As$_{4}$ is also studied as a function of magnetic field. A clear change in $H^\prime_{c2}$($T$)/$T_c$, where $H^\prime_{c2}$($T$) is d$H_{c2}$($T$)/d$T$, at $x$ $\sim$ 0.012 is observed and probably is related to change of the Fermi surface due to magnetic order. Coherence length and the London penetration depths are also calculated based on $H_{c1}$ and $H_{c2}$ data. Coherence lengths as the function of $x$ also shows changes near $x$ = 0.012, again consistent with Fermi surfaces changes associated with the magnetic ordering seen for higher $x$-values.
\end{abstract}

\maketitle

\section{introduction}

 The study of Fe-based superconductors has lead to extensive experimental interest and their variety offers the opportunity of understanding unconventional superconductivity in a broader sense. Fe-based superconductors families share similar crystal structures {\color{blue}\cite{Kamihara2008,Johnston2010,Paglione2010,Hosono2015,Yoshida2016,Meier2016,Meier2017,Hsu14262,Kawashima2016,Bao2018}} and phase diagrams {\color{blue}\cite{Canfield2009a,Ni2009,Ni20101,Canfield2010f,Stewart2011,Meier2018,Bohmer2017a,Gati2020, Mingyu01}}, which suggests a relationship between, or proximity of, superconducting and magnetic and/or nematic ordering or fluctuations. These relationships between superconductivity and magnetic as well as structural transitions and fluctuations are believed to be key to understand unconventional superconductivity {\color{blue}\cite{Mazin2010}}.

Though Fe-based superconductors have a moderate degree of structural diversity, three, main structural classes: $Ae$Fe$_2$As$_2$ ($Ae$=Alkaline Earth) (122) families {\color{blue}\cite{Canfield2009a,Ni2009,Ni20101,Johnston2010,Paglione2010,Canfield2010f,Stewart2011,Hosono2015}}, $AeA$Fe$_4$As$_4$ (A=Alkaline Metal) (1144) family {\color{blue}\cite{Yoshida2016,Meier2016,Meier2017,Meier2018,Kawashima2016,Bao2018}} and FeSe {\color{blue}\cite{Hsu14262,Bohmer2017a}} provide a microcosm of many key questions at hand. Compared with the interplay between stripe-like antiferromagnetic order, nematicity, and superconductivity in the doped 122-systems at ambient pressure {\color{blue}\cite{Canfield2010f,Paglione2010}}, the electron-doped CaK(Fe$_{1-x}T_x$)$_4$As$_4$, $T$ = Ni and Co system has hedgehog-spin-vortex-crystal (h-SVC) type antiferromagnetic (AFM) order and superconductivity interacting with each other without any structural phase transition{\color{blue}\cite{Meier2018,Meier2019T}}. 

On one hand, since the phase diagrams of Co and Ni substitutions of CaKFe$_{4}$As$_{4}$ scaled almost exactly as a function of band filling change, the comparison between CaKFe$_4$As$_4$ and Ba$_{0.5}$K$_{0.5}$Fe$_2$As$_2$ based on their similar, nominal electron counts, seems justified. One the other hand, given that CaK(Fe$_{1-x}T_x$)$_4$As$_4$ allows for the study of how nominal hole-doping with Mn and Cr can affect the superconducting and magnetic properties of this system, it is very important to see how their $T$-$x$ phase diagrams compare with each other as well as those for $T$ = Co and Ni.


We have recently found that for CaK(Fe$_{1-x}T_{x}$)$_4$As$_4$, $T$ = Mn, Mn is a far more local-moment-like impurity than $T$ = Co or Ni are. We also found that the substitution level of CaK(Fe$_{1-x}$Mn$_{x}$)$_{4}$As$_{4}$ can only go to up to $x$ = 0.036. Beyond that level, 1144 phase is not stabilized with the similar synthesis condition. This limited the exploration of hole -doped 1144 phase diagram and the evolution of h-SVC type antiferromagnetic transition. Cr offers twice the amount of nominal hole-doping per $x$ and, like Mn, can sometimes manifest local-moment-like properties in intermetallic samples. As such, Cr substitution offers a great of opportunity to further our understanding of the behavior of h-SVC type antiferromagnetism in the 1144 system.

In this paper, we detail the synthesis and characterization of CaK(Fe$_{1-x}$Cr$_x$)$_4$As$_4$ single crystals. A temperature-composition ($T$-$x$) phase diagram is constructed by elemental analysis, magnetic and transport measurements. In addition to creating the $T$-$x$ phase diagram, coherence lengths and the London penetration depths are also calculated based on $H_{c1}$ and $H_{c2}$ data obtained from measurements. The data for Cr-substituted 1144 are added on the $\lambda ^{-2}$ versus $\sigma$ $T_c^2$ plot and compared with the Mn substitution. Finally, temperature vs change of electron count, |$\Delta e^-$|, phase diagram for CaK(Fe$_{1-x}T_{x}$)$_{4}$As$_{4}$ single crystals, $T$ = Cr, Mn, Ni and Co, is also presented and discussed. By comparing all four $T$ = Cr, Mn, Ni and Co substitutions we find that whereas for $T$ = Ni and Co CaK(Fe$_{1-x}T_x$)$_4$As$_4$ the temperature -substitution phase diagrams scale with additional electrons (in much the same way that the Ba(Fe$_{1-x}TM_x$)$_2$As$_2$ phase diagrams do for $T$ = Ni and Co), for $T$ = Cr and Mn the temperature substitution phase diagrams are essentially identical when plotted more simply as $T$-$x$ diagrams, suggesting that for Cr and Mn there may be other variables or mechanisms at play.

\section{Crystal Growth and Experimental Method}

Single crystalline CaK(Fe$_{1-x}$Cr$_{x}$)$_{4}$As$_{4}$ samples were grown by high-temperature solution growth {\color{blue}\cite{Canfield2020}} out of FeAs flux in the manner similar to CaK(Fe$_{1-x}$Mn$_{x}$)$_{4}$As$_{4}$ {\color{blue}\cite{Mingyu01}}. Lumps of potassium metal (Alfa Aesar 99.95\%), distilled calcium metal pieces (Ames Laboratory, Materials Preparation Center (MPC 99.9\%) and Fe$_{0.512}$As$_{0.488}$ and Cr$_{0.512}$As$_{0.488}$ precursor powders were loaded into a 1.7 ml fritted alumina Canfield Crucible Set {\color{blue}\cite{Canfield2016a}} (LSP Industrial Ceramics, Inc.) in an argon filled glove-box. The ratio of K:Ca:Fe$_{0.512}$As$_{0.488}$ and Cr$_{0.512}$As$_{0.488}$ was 1.2:0.8:20. A 1.3~cm outer diameter and 6.4~cm long tantalum tube which was used to protect the silica ampoule from reactive vapors was welded with the crucible set in partial argon atmosphere inside. The sealed Ta tube was then itself sealed into a silica ampoule and the ampoule was placed inside a box furnace. The furnace was held for 2 hours at 650~\textcelsius\ before increasing to 1180~\textcelsius\ and held there for 5 hours to make sure the precursor was fully melted. The furnace was then fast cooled from 1180~\textcelsius\ to 980~\textcelsius\ in 1.5 hours. Crystals were grown during a slow cool-down from 980~\textcelsius\ to 915~\textcelsius\ over 100-150 hours dependent on substitution level. After 1-2 hours at 915~\textcelsius\, the ampoule was inverted into a centrifuge and spun to separate the remaining liquid from the grown crystals. Metallic, plate-like, crystals were obtained.~The average size and thickness decreased by factor 2-4 as $x$ is increased. The largest crystal is about centimeter size as shown in figure \ref{figure1}

Single crystals of CaK(Fe$_{1-x}$Cr$_{x}$)$_{4}$As$_{4}$ are soft and malleable as CaKFe$_{4}$As$_{4}$ and are difficult to grind for powder X-ray diffraction measurements. Diffraction measurements were carried out on single crystal samples, which were cleaved along the (001) plane, using a Rigaku MiniFlex II powder diffactometer in Bragg-Brentano geometry with Cu K$\alpha$ radiation ($\lambda$ = 1.5406 \AA{}) {\color{blue}\cite{Jesche2016}}.

The Cr substitution levels ($x$) of the CaK(Fe$_{1-x}$Cr$_{x}$)$_{4}$As$_{4}$ crystals were determined by energy dispersive spectroscopy (EDS) quantitative chemical analysis using an EDS detector (Thermo NORAN Microanalysis System, model C10001) attached to a JEOL scanning-electron microscope. The compositions of platelike crystals were measured at three separate positions on each crystal's face (parallel to the crystallographic \textit{ab}-plane) after cleaving them. An acceleration voltage of 16~kV, working distance of 10~mm and take off angle of 35$^{\circ}$ were used for measuring all standards and crystals with unknown composition. Pure CaKFe$_4$As$_4$ was used as a standard for Ca, K, Fe and As quantification. LaCrGe$_3$ and YCr$_6$Ge$_6$ were used as standards for Cr, both leading to consistent results without significant difference within the experimental error ($\sim$ 0.001). The spectra were fitted using NIST-DTSA II Microscopium 2020-06-26 software\cite{Newbury2014}. Different measurements on the same sample reveal good homogeneity in each crystal and the average compositions and error bars were obtained from these data, accounting for both inhomogeneity and goodness of fit of each spectra.

Temperature- and magnetic-field-dependent magnetization and resistance measurements were carried out by using Quantum Design (QD), Magnetic Property Measurement Systems (MPMS and MPMS3) and Physical Property Measurement Systems (PPMS). Temperature- and magnetic-field-dependent magnetization measurements were taken for \textit{H}$\parallel$\textit{ab} by placing the plate-like sample between two collapsed plastic straws with the third, uncollapsed, straw providing support as a sheath on the outside or by using of a quartz sample holder. The single crystal samples of CaK(Fe$_{1-x}$Cr$_{x}$)$_{4}$As$_{4}$ measured in the MPMS and MPMS3 have plate-like morphology with length and width from 3 mm to 10 mm and thickness (\textit{c} axis) 50 - 200 $\mu$m. The approximate effective demagnetizing factor $N$ ranges from 0.007 to 0.077 with magnetic field applied parallel to the crystallographic \textit{ab} plane{\color{blue}\cite{Prozorov2018}}. AC electrical resistance measurements were performed in a standard four-contact geometry using the ACT option of the PPMS, with a 3 mA excitation and a frequency of 17 Hz. 50$\mu$m diameter Pt wires were bonded to the samples with silver paint (DuPont 4929N) with contact resistance values of about 2-3 Ohms. The magnetic field, up to 90 kOe, was applied along \textit{c} or \textit{ab} directions, perpendicular to the current, with the current flowing in the \textit{ab} plane in both cases.


Contacts for inter-plane resistivity measurements were soldered using tin.  
The top and bottom surfaces of the samples were covered with Sn solder {\color{blue}\cite{Tanatar2010a,MAKARIY2013}} and 50 $\mu$m silver wires were attached to enable measurements in a pseudo-four-probe configuration.
Soldering produced contacts with resistance typically in the 10 $\mu \Omega$ range. Inter-plane resistivity was measured using a two-probe technique with currents in 1 to 10 mA range (depending on sample resistance which is typically 1 m$\Omega$). A four-probe scheme was used down to the sample to measure series connected sample, $R_s$, and contact, $R_c$ resistance. Taking into account that $R_s \gg R_c$, contact resistance represents a minor correction of the order of 1 to 5\%. 
The details of the measurement procedure can be found in Refs.~{\color{blue}\cite{Tanatar2009, Tanatar2009a, Tanatar2010}}. The results of the measurements are in good agreement with similar measurements on pure CaKFe$_4$As$_4$ \cite{Meier2016}. 

Measurements with current along the $c-$axis suffer strongly from inter-layer connectivity due to the micacious nature of single  crystals. To ascertain reproducibility, we performed measurements of $\rho_c$ on two to five samples and obtained qualitatively similar temperature dependencies of the electrical resistivity, as represented by the ratio of resistivities at room and low temperatures, $\rho _c (0)/\rho _c (300)$. The resistivity $\rho_c(300K)$ was in the range 1- 2 m$\Omega cm$, corresponding to an anisotropy ratio $\rho_c/\rho_a\approx$ 3 to 6 at 300~K.

\section{C\lowercase{a}K(F\lowercase{e}$_{1-x}$C\lowercase{r}$_{x}$)$_{4}$A\lowercase{s}$_{4}$ Structure and Composition}

\begin{figure}
	\includegraphics[width=1.5\columnwidth]{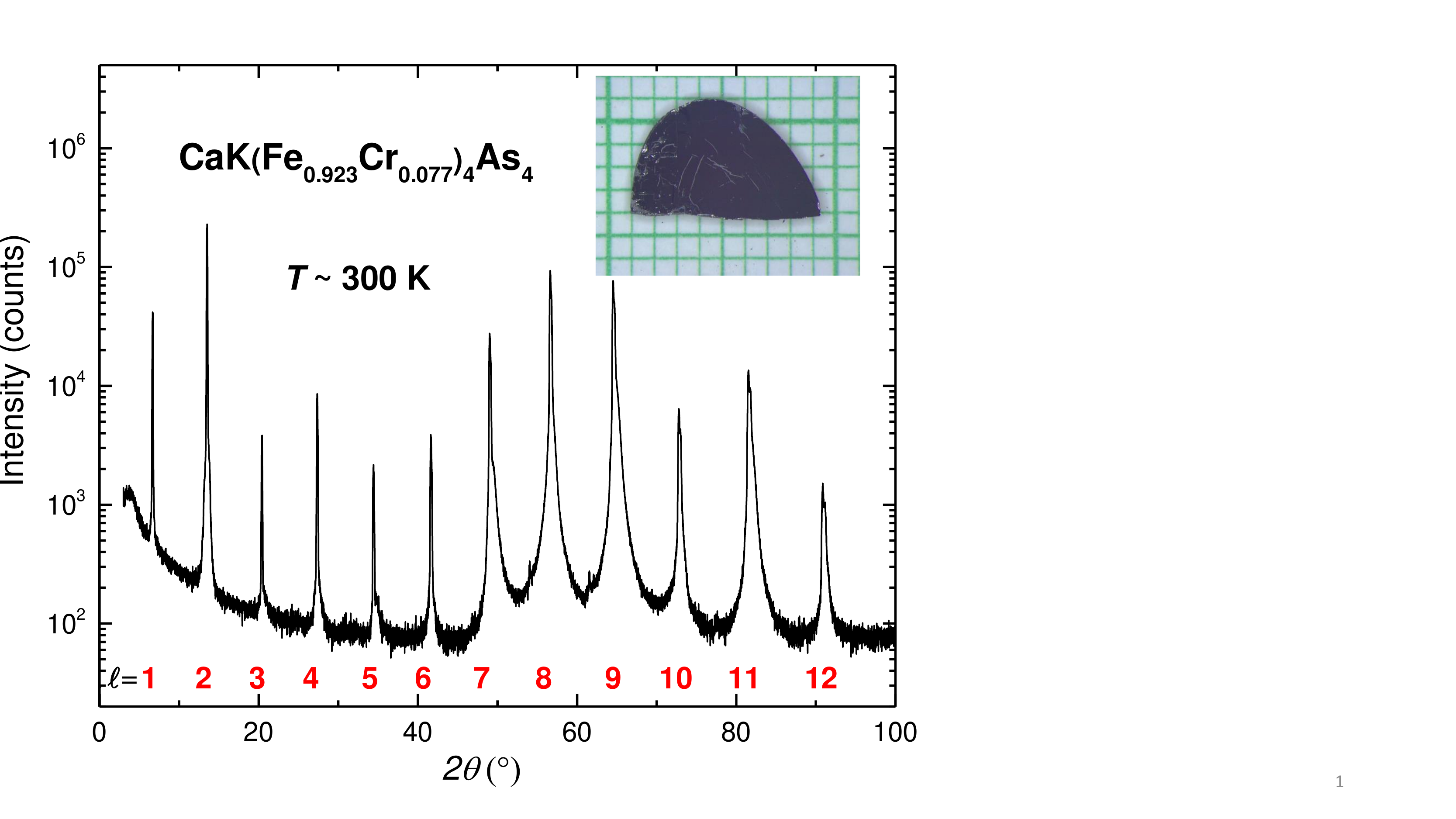}	
	\caption{ X-ray diffraction data of CaK(Fe$_{0.923}$Cr$_{0.077}$)$_{4}$As$_{4}$ showing ($00l$) diffraction peaks from in-lab diffraction measurements on a single-crystalline plate plotted on a semi-log scale. The value of $l$ is shown in red under each peak. The inset shows the picture of  CaK(Fe$_{0.988}$Cr$_{0.012}$)$_{4}$As$_{4}$ single crystal. Note that $l$ = odd ($00l$) lines are evidence of the ordered CaKFe$_{4}$As$_{4}$ structure {\color{blue}\cite{Yoshida2016,Meier2016}}.\label{figure1}}
\end{figure}

\begin{figure}
	\centering
	\begin{minipage}{0.44\textwidth}
		\centering
		\includegraphics[width=1.5\columnwidth]{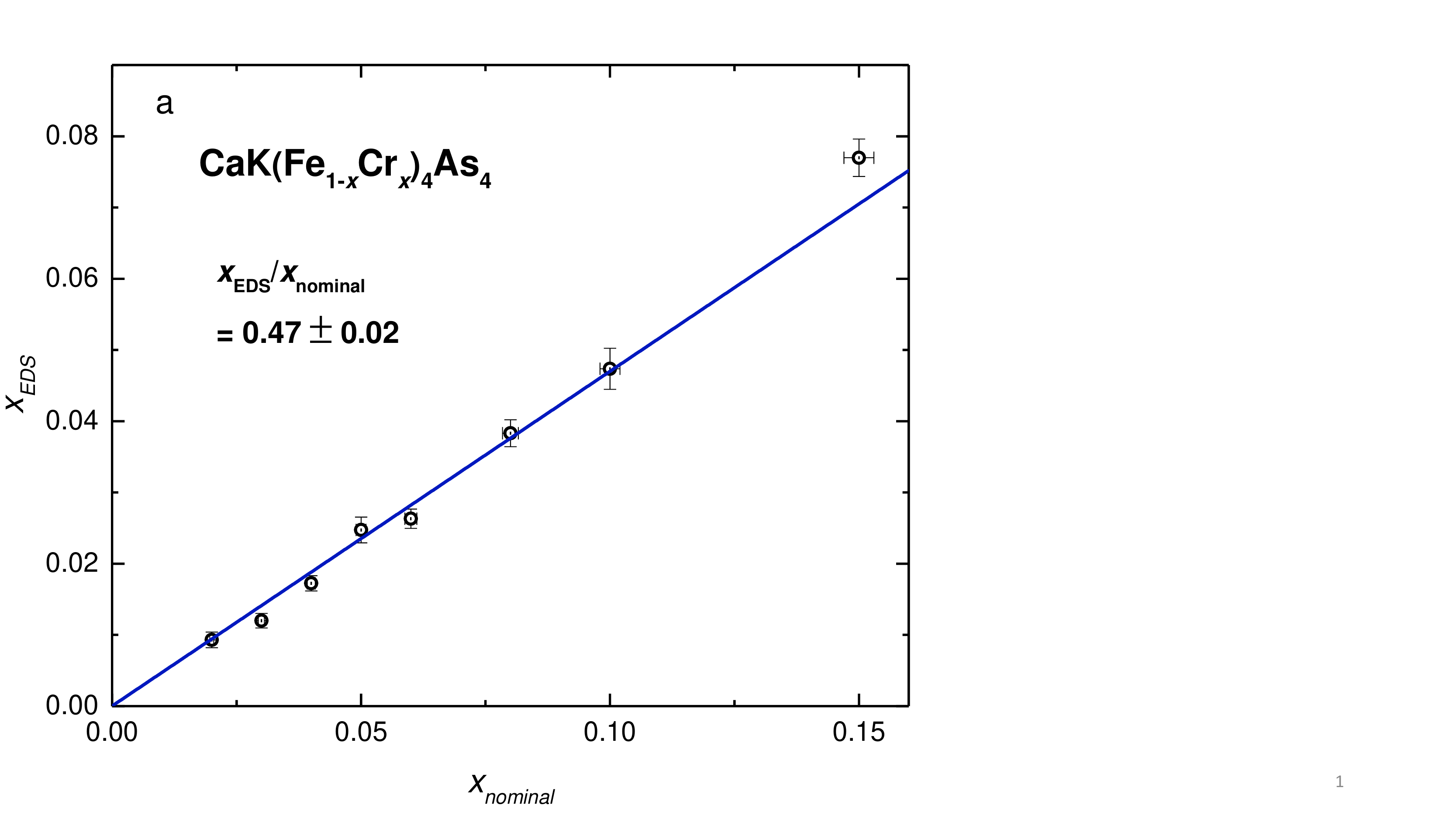}		
	\end{minipage}\hfill
	\centering
	\begin{minipage}{0.44\textwidth}
		\centering
		\includegraphics[width=1.5\columnwidth]{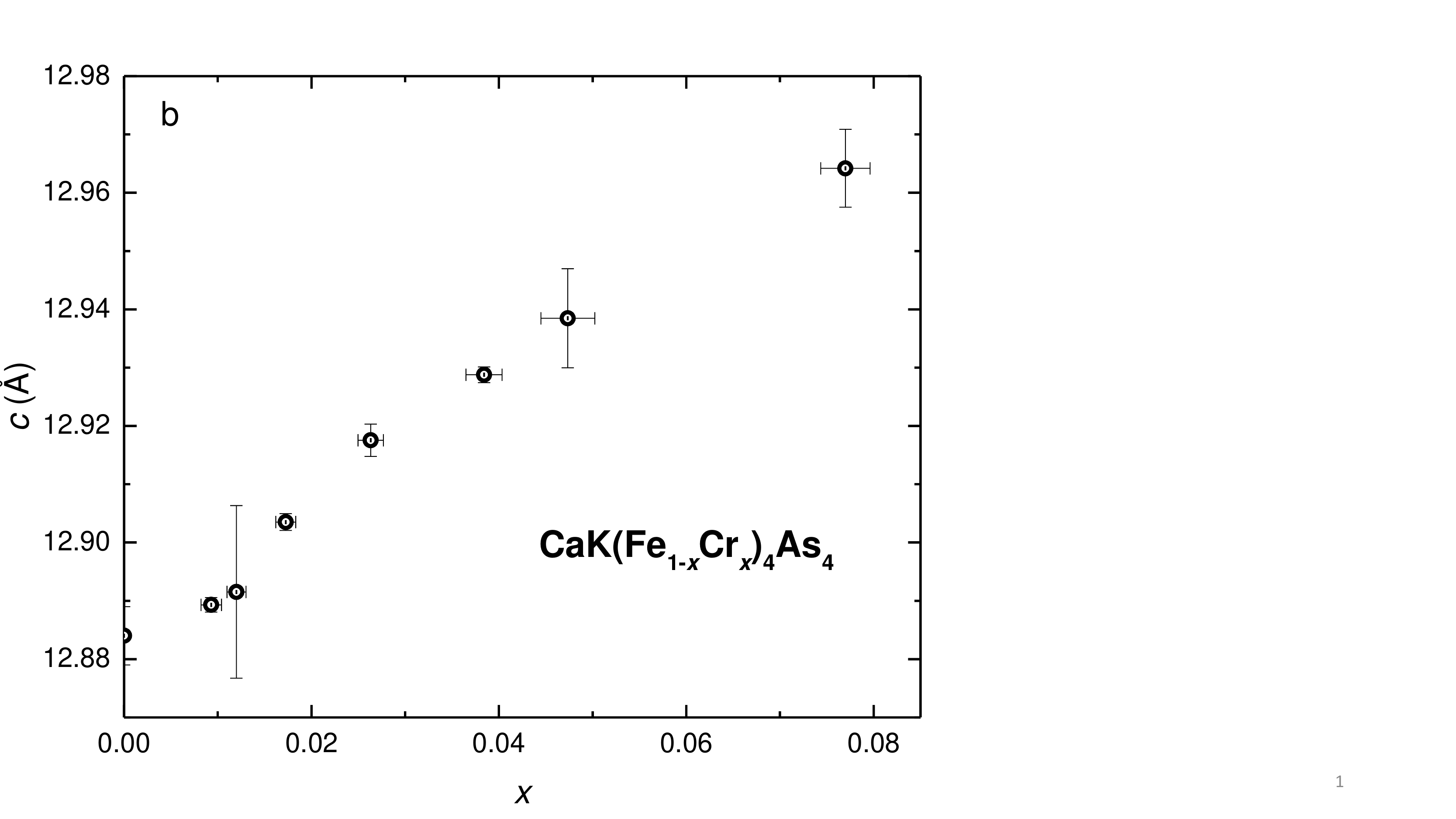}		
	\end{minipage}\hfill
	
	\caption{(a)EDS measured Cr concentration vs nominal Cr concentration for the CaK(Fe$_{1-x}$Cr$_{x}$)$_{4}$As$_{4}$ series. The line is the linear fit of the data with fixed intercept to 0. (b) Plot of $c$ lattice parameters as a function of substitution level $x$. $c$ lattice parameters are calculated by the single-crystalline plate X-ray diffraction plot {\color{blue}\cite{Jesche2016}}.}\label{figure5}
\end{figure}

 Figure \ref{figure1} presents single crystal diffraction data of CaK(Fe$_{1-x}$Cr$_{x}$)$_{4}$As$_{4}$ with $x_{EDS}=0.077$ which is the largest substitution level obtained.~Attempts to grow crystals with $x_{EDS}$ > 0.077 failed to yield mm-sized or larger samples that could be identified as Cr-doped 1144. From the figure, we can see that all ($00l$), $l\leq12$, are detected. The $h+k+l$ = odd peaks which are forbidden for the \textit{I}4/mmm  structure {\color{blue}\cite{Yoshida2016}} can be clearly found. This indicates that sample has the anticipated \textit{P}4/mmm structure associated with the CaKFe$_4$As$_4$ structure {\color{blue}\cite{Yoshida2016,Meier2018,Meier2016}}.

 The Cr substitution, $x$$_{EDS}$, determined by EDS is shown in figure \ref{figure5}a~for different crystals as a function of the nominal Cr fraction, $x$$_{nominal}$, that was originally used for the growth. Error bars account for both possible inhomogeneity of substitution and goodness of fit of each EDS spectra. A clear correlation can be seen between the nominal and the measured substitution levels, with a proportionality factor of 0.47 $\pm$ 0.02. For comparison, the ratio of measured to nominal Mn, Ni and Co fraction in the corresponding CaK(Fe$_{1-x}$\textit{T}$_x$)$_{4}$As$_{4}$ are 0.60, 0.64 and 0.79 respectively {\color{blue}\cite{Meier2018,Mingyu01}}. From this point onward, when substitution level $x$ is referred to, it will be the EDS value of $x$. Figure \ref{figure5}b presents $c$ lattice parameters as the function of $x$. $c$ lattice parameter monotonically increases as Cr substitution level increase, which is consist with the larger radius of Cr than Fe. $c$ lattice parameter values are calculated by the single-crystalline plate X-ray diffraction plot {\color{blue}\cite{Jesche2016}}. In Mn-1144 (CaK(Fe$_{1-x}$Mn$_{x}$)$_{4}$As$_{4}$), the evolution of the $c$ lattice parameter is difficult to determine due to the small difference of radius between Fe and Mn and low substitution levels. The highest $x$ is 0.036 in Mn-1144 and it's smaller than, 0.039, lowest substitution of Co-1144. 
 
\section{Data analysis and Phase Diagram}
 
\begin{figure}
	\includegraphics[width=1.5\columnwidth]{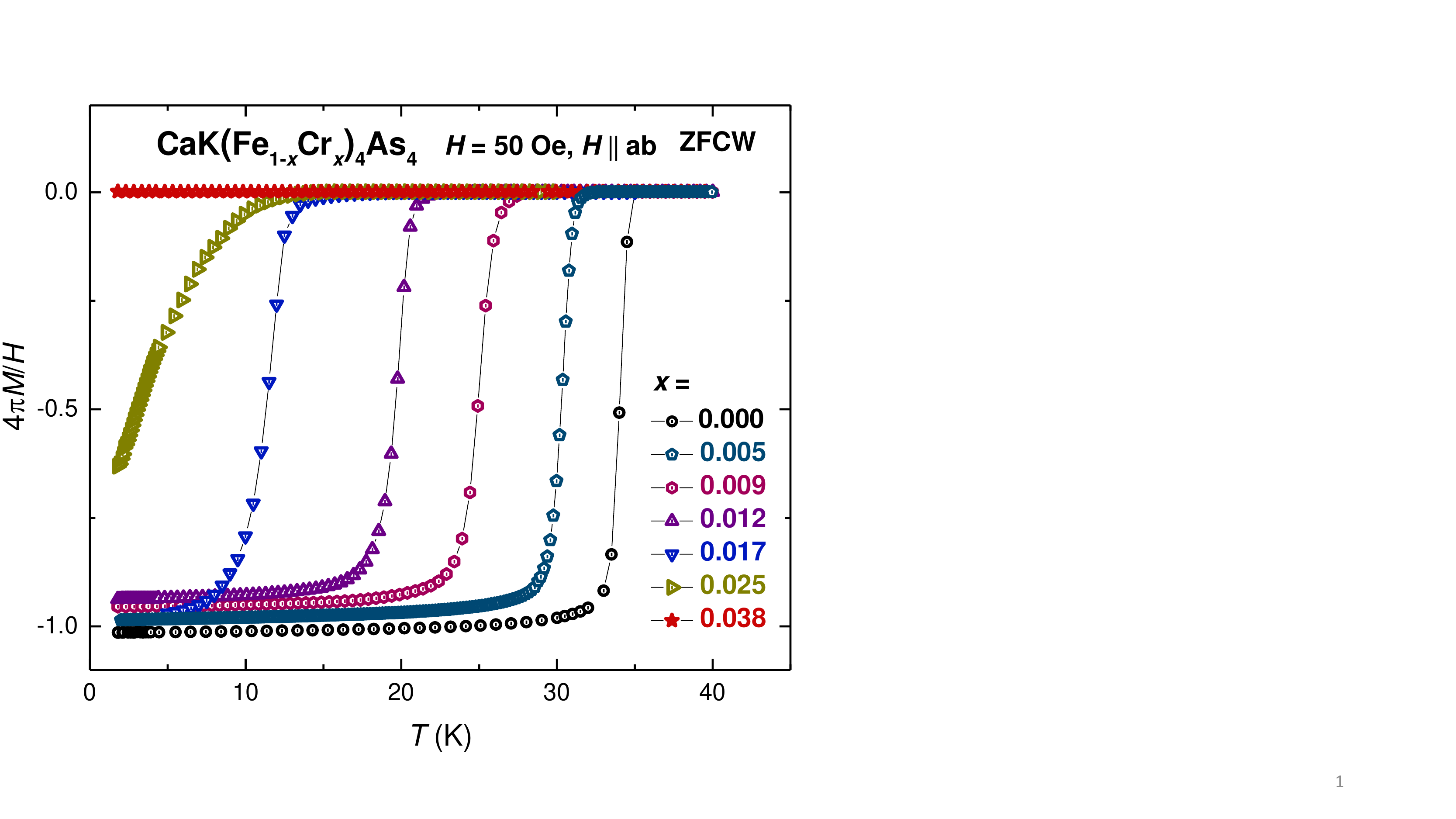}	
	\caption{Zero-field-cooled-warming (ZFCW) low temperature magnetization as a function of temperature for CaK(Fe$_{1-x}$Cr$_{x}$)$_{4}$As$_{4}$ single crystals with a field of 50~Oe applied parallel to the crystallographic \textit{ab} plane. $M$ is the volumetric magnetic moment with cgs unit emu cm$^{-3}$ or Oe. \label{figure2}}
\end{figure}
\begin{figure}
	\includegraphics[width=1.5\columnwidth]{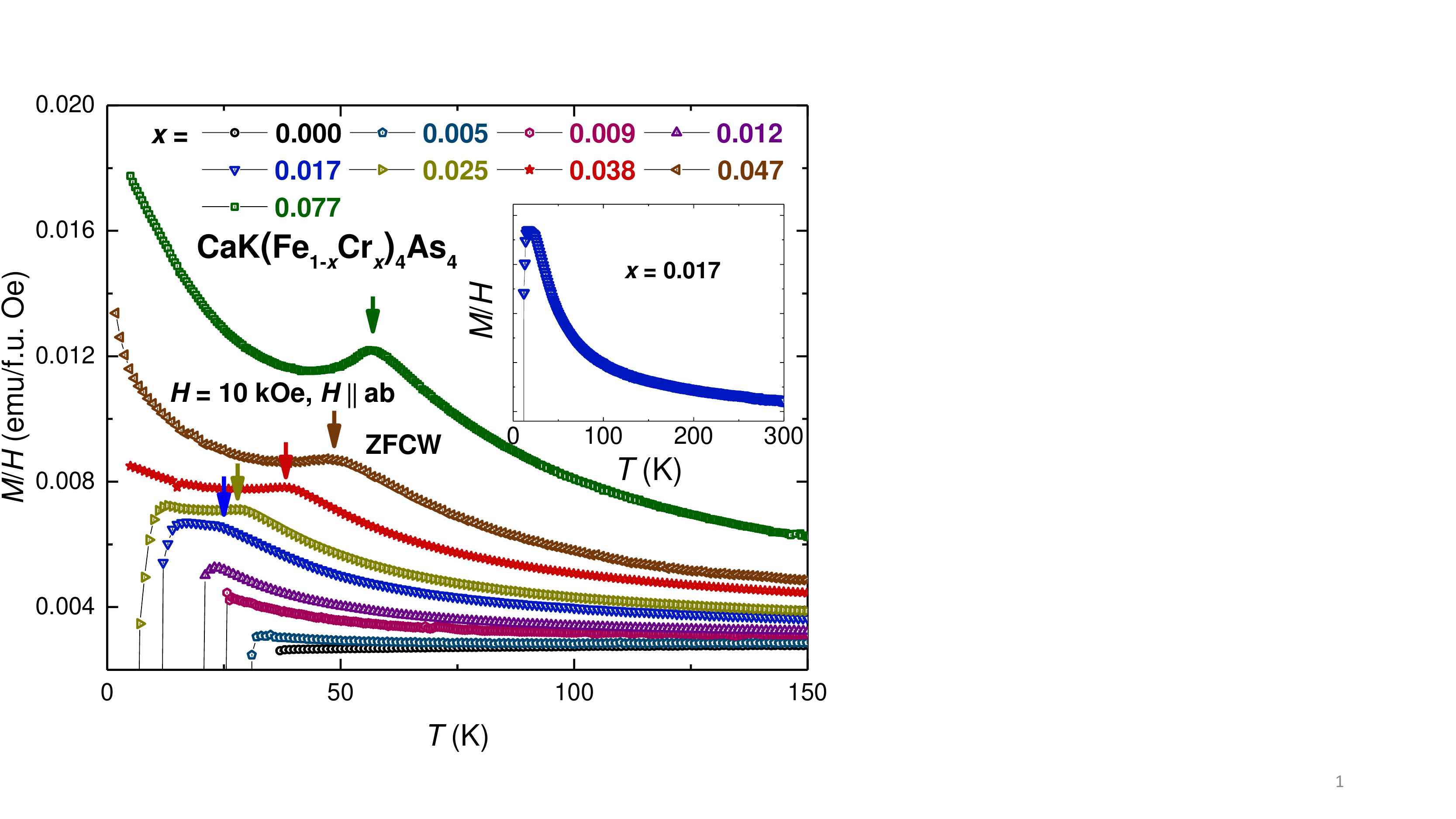}		
	\caption{Low temperature magnetization divided by applied field as a function of temperature for CaK(Fe$_{1-x}$Cr$_{x}$)$_{4}$As$_{4}$ single crystals with a field of 10~kOe applied parallel to the crystallographic \textit{ab} plane. The inset shows the CaK(Fe$_{0.983}$Cr$_{0.017}$)$_{4}$As$_{4}$ single crystal magnetization for 5 K < $T$ < 300 K. Small vertical arrows indicate the location of $T^*$, see Appendix for the criterion. \label{figure3}}
\end{figure}
\begin{figure}
	\includegraphics[width=1.5\columnwidth]{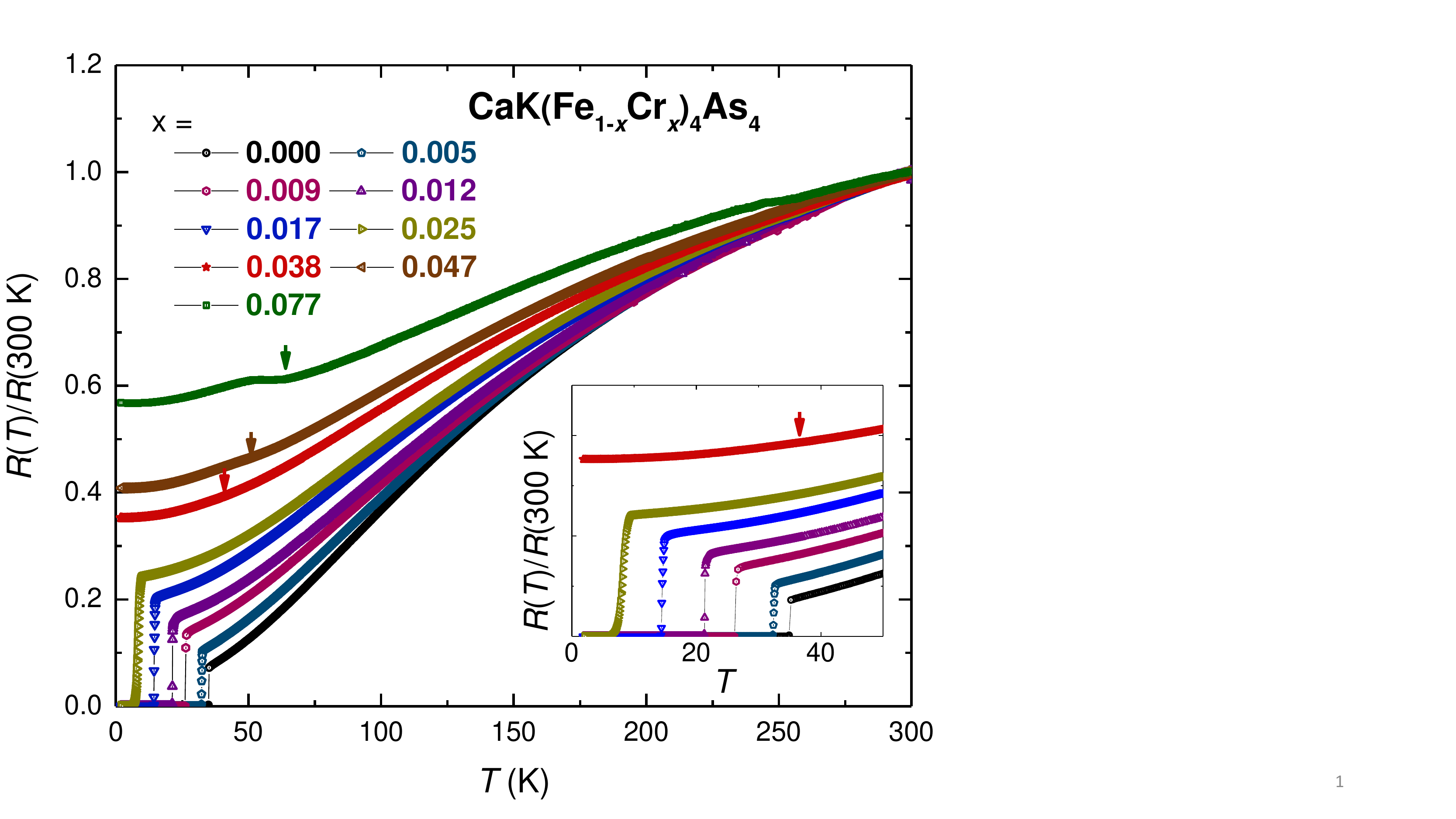}		
	\caption{Temperature dependence of normalized resistance, $\textit{R}(\textit{T})/\textit{R}$(300~K), of CaK(Fe$_{1-x}$Cr$_{x}$)$_{4}$As$_{4}$ single crystals showing the suppression of the superconducting transition $T_c$ and the appearance and evolution of a kink-like feature, marked with arrows. The criterion used to determine $T^*$ from this kink-feature is outlined and discussed in the appendix. \label{figure4}}
\end{figure}

Figure \ref{figure2} shows the low temperature (1.8~K - 45~K), zero-field-cooled-warming (ZFCW) magnetization for CaK(Fe$_{1-x}$Cr$_{x}$)$_{4}$As$_{4}$ single crystals for $H_{|| ab}$ = 50 Oe (ZFCW magnetization and field-cooled (FC) data for an $x$ = 0.017 sample can be found in figure \ref{50OeFCZFCz} in the Appendix). $M$ is the volumetric magnetization in this figure and is calculated by using the density of CaKFe$_4$As$_4$, which is determined to be 5.22~g/cm$^3$ from the lattice parameters at room temperature {\color{blue}\cite{Yoshida2016}}. A magnetic field of 50~Oe was applied parallel to \textit{ab} plane (i.e. parallel to the surface of the plate-like crystal). The superconducting transitions ($T_c$) are clearly seen in this graph except for the substitution value \textit{x} = 0.038. As the value of the Cr substitution, $x$, increases, the superconducting transition temperature decreases. For $x=0.025$, a full magnetic shielding is not reached by 1.8 K.

Figure \ref{figure3} shows the low temperature (5~K - 150~K) $M$($T$)/$H$ data for CaK(Fe$_{1-x}$Cr$_{x}$)$_{4}$As$_{4}$ single crystals with 10~kOe field applied parallel to the crystallographic \textit{ab} plane. The appearance of first a Curie-Weiss tail and later a kink-like feature after adding Cr is similar to Mn substituted 1144. Kink-like features are found above 20~K for $x>0.012$, which very likely indicate antiferromagnetic transition. Similar kink-like features were correlated with AFM order in CaK(Fe$_{1-x}$Mn$_{x}$)$_{4}$As$_{4}$ {\color{blue}\cite{Mingyu01,Wilde2023}}.~The inset shows $M$($T$)/$H$ of a CaK(Fe$_{0.983}$Cr$_{0.017}$)$_{4}$As$_{4}$ single crystal over a wider temperature range. As Cr is added the Curie-tail-like feature grows. The $M$($T$) data above transitions can be fitted by a C/($T$+$\theta$) + $\chi_{0}$ function as long as Cr doping levels is larger than 0.005 ($x$ > 0.005). The effective moment versus $x$ data is shown in figure \ref{figure 13} in the Appendix; $\mu_{eff}$ calculated per Cr, is found to be $\sim$ 4 $\mu$B. For $x$ > 0.012, a kink-like feature can be seen at a temperature $T^*$. As $x$ increases from 0.017 to 0.077, the temperature $T^*$  increases from $\sim$ 20 K to $\sim$ 60 K. The criterion for determining $T^*$ and more discussion about the Curie-tail are shown in the Appendix.

Figure \ref{figure4} presents the temperature dependent, normalized, electrical resistance of CaK(Fe$_{1-x}$Cr$_{x}$)$_{4}$As$_{4}$ single crystals. RRR (the ratio of 300 K and low temperature resistance just above $T_c$) decreases as Cr substitution increases, which is consistent with the disorder increasing. The superconducting transition temperatures decrease as Cr is added to the system. When \textit{x} = 0.038, there is no signature of a superconducting transition detectable above 1.8 K. With increasing Cr content, a kink appears for $x$ > 0.025 and rises to about 60~K for $x$ = 0.077 and features become more clearly resolved with increasing substitution. A similar feature also appeared in Mn, Ni and Co-substituted CaKFe$_{4}$As$_{4}$ electrical resistance measurements {\color{blue}\cite{Meier2018,Mingyu01}}. The criterion for determining the transition temperature, $T^*$, associated with this kink is shown in the Appendix in figure \ref{figure92}c where $R(T)$ and d$R(T)$/d$T$ are both shown.

\begin{figure}
	\includegraphics[width=1.5\columnwidth]{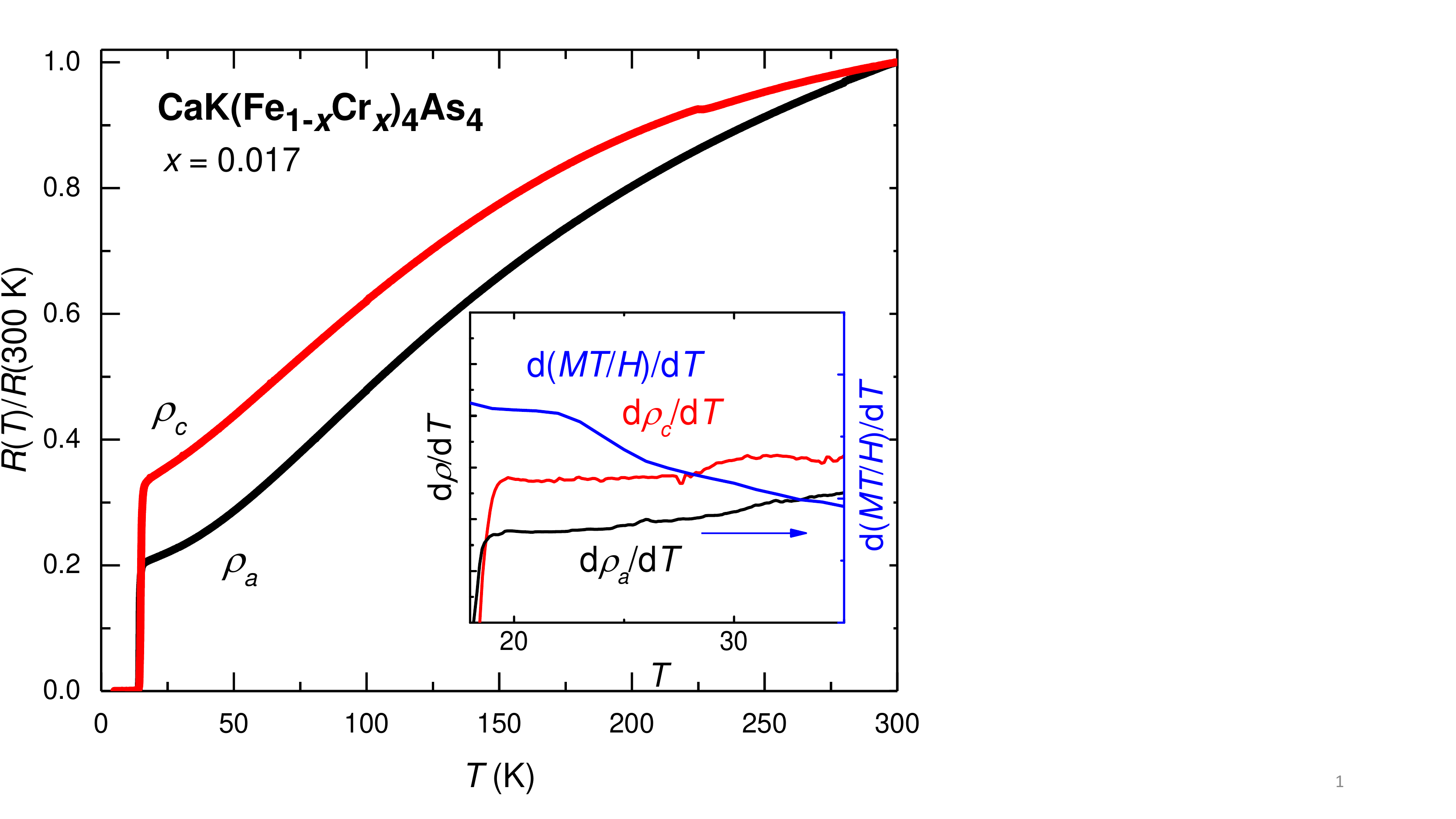}		
	\caption{Temperature dependence of normalized resistance, $\textit{R}(\textit{T})/\textit{R}$(300~K), of CaK(Fe$_{1-x}$Cr$_{x}$)$_{4}$As$_{4}$ single crystals with $x=$0.017 for electrical currents along $a-$axis (black) and along $c$-axis (red). Inset shows resistivity derivatives plotted against d($MT$/$H$)/d$T$ (blue). The onset of the feature in the d($MT$/$H$)/d$T$ curve at $\sim$25~K is accompanied by flattening in the $c$-axis resistivity derivative, but without a clear anomaly. No features are observed in $d(R/R(300 K))/dT$ curve for in-plane current  (black line in the inset.) 
		\label{resfigrhoc1}}
\end{figure}

\begin{figure}
	\includegraphics[width=1.5\columnwidth]{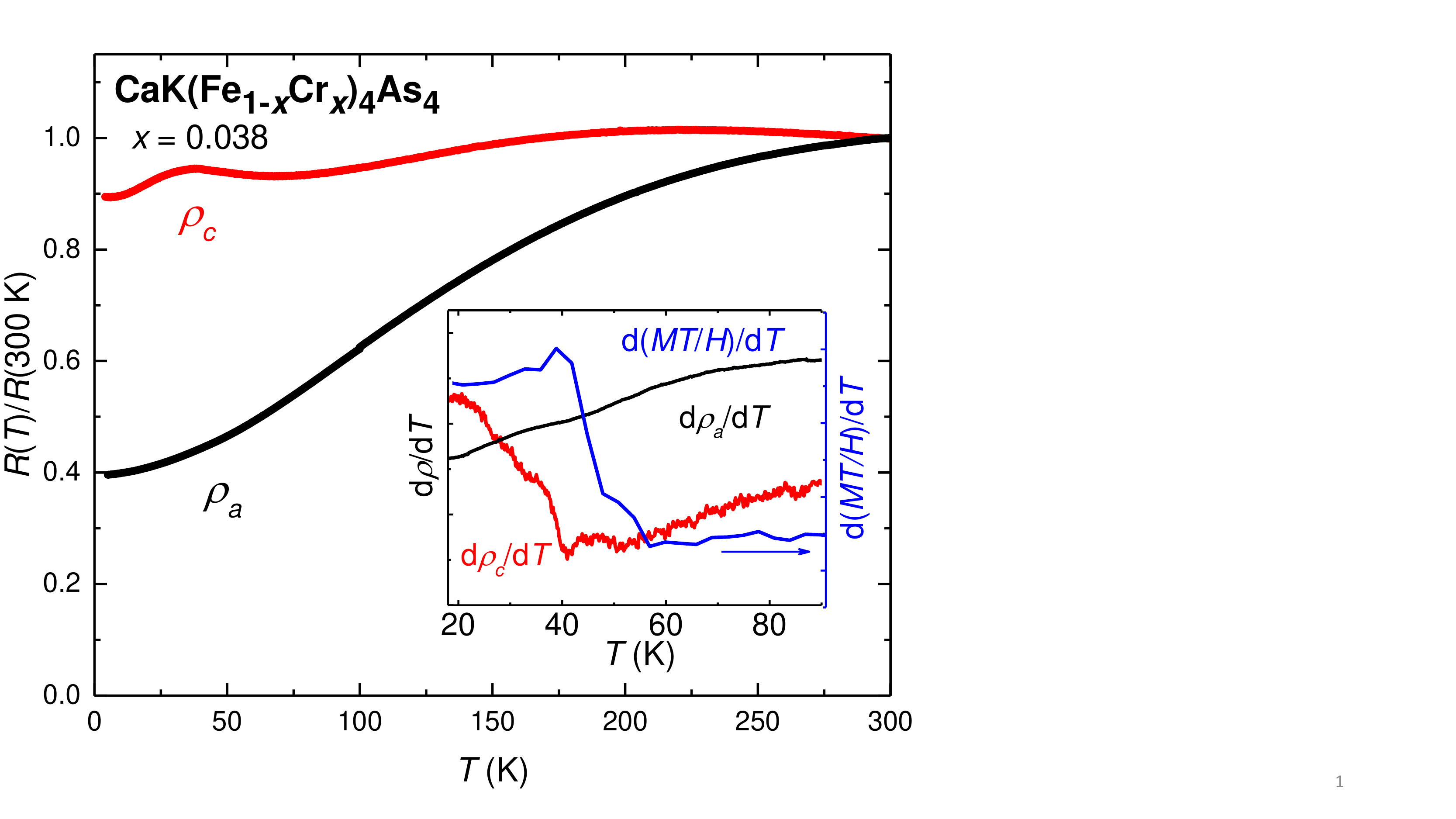}		
	\caption{Temperature dependence of normalized resistance, $\textit{R}(\textit{T})/\textit{R}$(300~K), of CaK(Fe$_{1-x}$Cr$_{x}$)$_{4}$As$_{4}$ single crystals with $x$ = 0.038 for electrical currents along $a$-axis (black) and along $c$-axis (red). Inset shows resistivity derivatives plotted against d($MT$/$H$)/d$T$ (blue, right axis). The peak in the d($MT$/$H$)/d$T$ curve at $\sim$ 42~K is accompanied by a clear feature in the $c$-axis resistivity derivative, but only with a very broad anomaly for the in-plane transport. 
		\label{resfigrhoc2}}
\end{figure}

\begin{figure}
	\includegraphics[width=1.5\columnwidth]{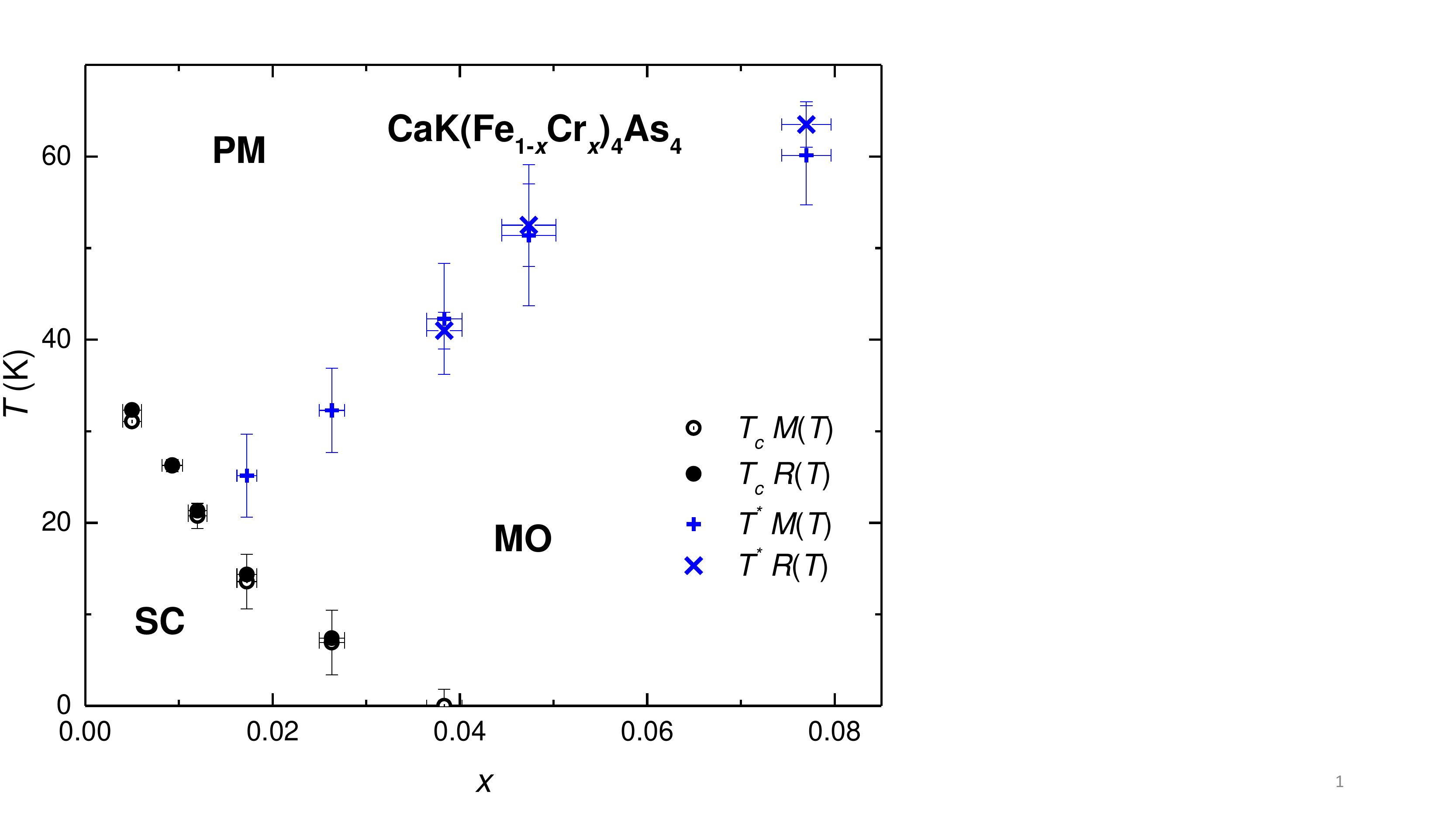}		
	\caption{Temperature - composition phase diagram of CaK(Fe$_{1-x}$Cr$_{x}$)$_{4}$As$_{4}$ single crystals as determined from resistance ($R$($T$)) and magnetization ($M$($T$)) measurements. The circular symbols denote the $T_{c}$ phase line and the cross-like symbols denote the $T^{*}$ phase line, most likely associated with antiferromagnetic order. Superconducting (SC), magnetically ordered (MO) and paramagnetic (PM) regions are denoted. Details of how the MO line extends into the SC state are not addressed in this phase diagram.    \label{figure11}}
\end{figure}

Figures \ref{resfigrhoc1} and \ref{resfigrhoc2} compare the normalized electrical resistivity of CaK(Fe$_{1-x}$Cr$_{x}$)$_{4}$As$_{4}$ single crystals for electrical currents along $a$-axis in the tetragonal plane (black lines in the main panels) with those along the tetragonal $c$-axis (red curves in the main panels). Samples with $x$ = 0.017, figure~\ref{resfigrhoc1}, are in the range of SC and AFM coexistence; samples with $x$ = 0.038, figure~\ref{resfigrhoc2}, are in the range where superconductivity is suppressed (see figure \ref{figure11}, below). The inter-plane resistivity of the samples with $x$ = 0.017 shows a broad cross-over close to room temperature,  much milder feature is found in in-plane transport. This is very similar to the results on the parent compound $x$ = 0 {\color{blue}\cite{Meier2016}}. For samples with $x$ = 0.038, figure~\ref{resfigrhoc2}, the inter-plane resistivity (red curve, main panel)  reveals clearly non-monotonic dependence. The cross-over transforms into a clear maximum above 200~K, followed by a second maximum centered at about 35~K, close to the temperature of long-range magnetic ordering. 

Insets in the figures compare the derivatives of the normalised resistivities for two current directions with d($MT$/$H$)/d$T$ (blue lines, right scale) {\color{blue}\cite{Fisher1962}}. For sample $x$ = 0.017 in figure~\ref{resfigrhoc1}, no clear features are observed in the resistivity derivatives, however some flattening is  observed for $c$-axis resistivity. For samples with $x$ = 0.038, the peak in  d($MT$/$H$)/d$T$ at $\sim$ 42~K is in good agreement with the feature in the derivative of inter-plane resistivity. A feature in the derivative of in-plane resistivity is notably less pronounced (black curve in inset in figure~\ref{resfigrhoc2}).

In some cases the features associated with magnetic ordering in the FeAs-based superconductors are clearer for current flow along the $c$-axis as opposed to current flow in the basal $ab$-plane {\color{blue}\cite{Tanatar2010}}. This is believed to be due to an anternating arrangements of the magnetic moments along the $c$-axis direction, providing partial gapping of the Fermi surface affecting more strongly inter-plane transport. The clarity of features increases with $x$ in CaK(Fe$_{1-x}$Cr$_{x}$)$_{4}$As$_{4}$, making then clearly visible in the raw resistivity data for $x=$ 0.077 (see Appendix figure \ref{resfigrhoc3}). 


Figure \ref{figure11} summarizes the transition temperature results inferred from magnetization and resistance measurements, plots the superconducting and magnetic transitions as a function of substitution and constructs the $T$-$x$ phase diagram for the CaK(Fe$_{1-x}$Cr$_{x}$)$_{4}$As$_{4}$ system.~As depicted in this phase diagram, increasing Cr substitution (i) suppresses $T_c$ monotonically with it extrapolating to 0 K by $x$ $\sim$ 0.03 and (ii) stabilizes a new transition, presumably an antiferromagnetic one, for $x$ $\gtrsim$ 0.017 with the transition temperature rising from $\sim$ 20 K for $x$ = 0.017 to $\sim$ 60 K for $x$ = 0.077. Each phase line is made out of data points inferred from \textit{R}(\textit{T}) and \textit{M}(\textit{T}) measurements, illustrating the good agreement between our criteria for inferring $T_{c}$ and $T^{*}$ from magnetization and resistivity data. The CaK(Fe$_{1-x}$$T$$_{x}$)$_{4}$As$_{4}$ series, $T$ = Mn, Co and Ni, have qualitatively similar phase diagrams, with the quantitative differences being associated with the substitution levels necessary to induce the magnetic phase and to suppress superconductivity. We were not able to infer the behavior of $T^*$ once it drops below $T_c$, but if it is similar to other $T$ substitution{\color{blue}\cite{BudKo2018,Wilde2023}}, $T^{*}$ should be suppressed very fast in the superconductivity state. Further comparison of the CaK(Fe$_{1-x}$Cr$_{x}$)$_{4}$As$_{4}$ phase diagram to the phase diagrams of the CaK(Fe$_{1-x}$$T$$_{x}$)$_{4}$As$_{4}$ series will be made in the discussion section below.

\begin{figure}
	\includegraphics[width=1.5\columnwidth]{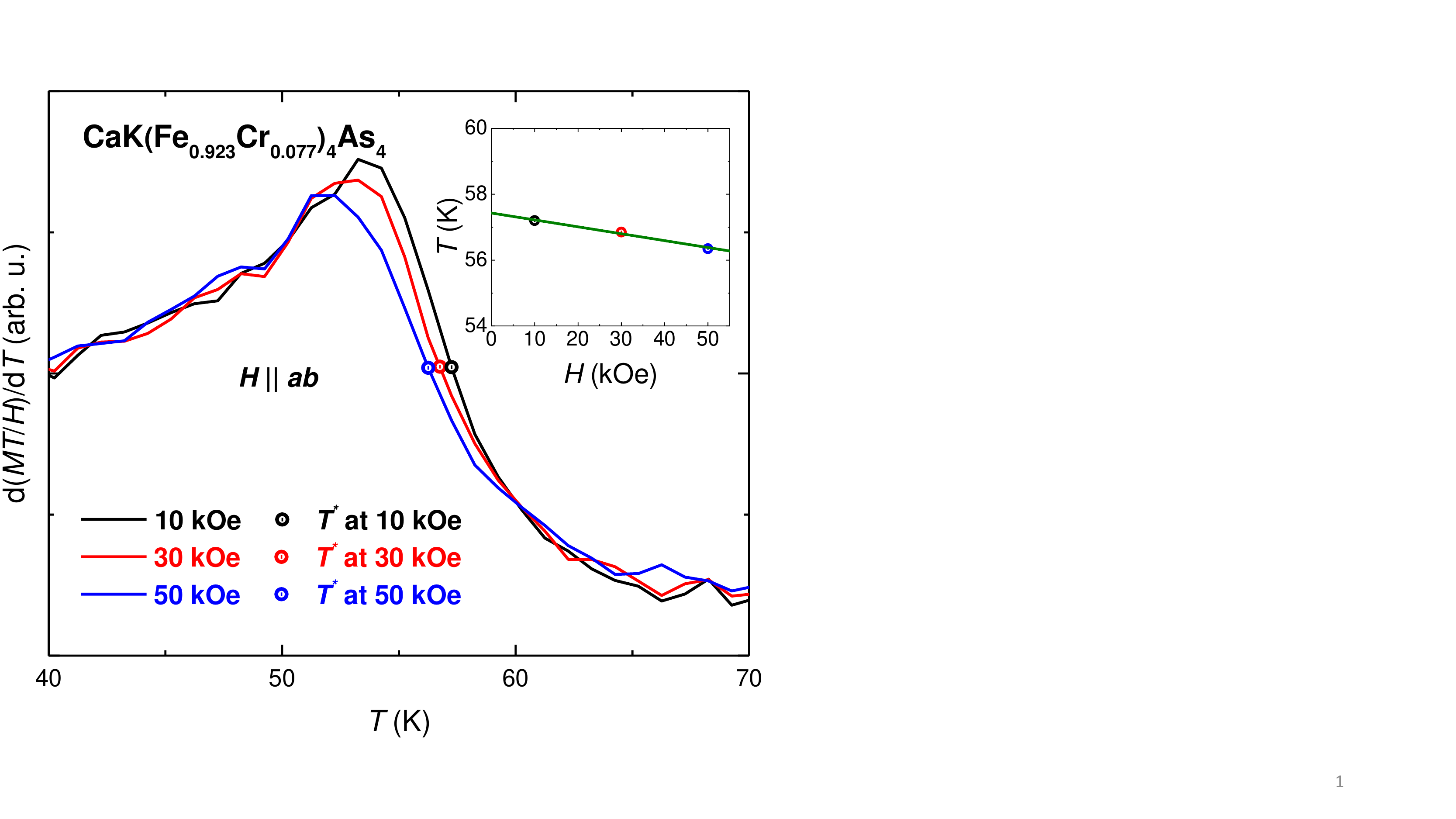}		
	\caption{d($MT/H$)/d$T$ vs. $T$ of CaK(Fe$_{0.923}$Cr$_{0.077}$)$_{4}$As$_{4}$ single crystal with 10 kOe, 30 kOe and 50 kOe applied parallel to the crystallographic \textit{ab} plane. $T^*$ at different field are shown as points on the figure. Inset shows transition temperature, $T^*$, inferred for different applied field values using the same criterion shown in appendix. The solid green line is linear fit to the data points, extrapolating to 54.7 K for $H$ = 0.   \label{figure15}}
\end{figure}

Given that the $R$($T$) data were taken in zero applied field whereas the $M/H$($T$) data shown in figure \ref{figure3} were taken in 10 kOe, it is prudent to examine the field dependence of transition associated with $T^*$. In figure \ref{figure15} we show the d($MT/H$)/d$T$ data {\color{blue}\cite{MFisher1962}} for the $x$ = 0.077 sample for $H$ || \textit{ab} = 10, 30 and 50 kOe. As is commonly seen for antiferromagnetic phase transition, increasing a magnetic field leads to a monotonic suppression of $T^{*}$. The inset to figure \ref{figure15} shows that the extrapolated, $H$ = 0, $T^*$ value would be 57.4 K as compared to the value of 57.2 K for 10 kOe. This further confirms that there should be (and is) good agreement between the $T^{*}$ values inferred from 10 kOe magnetization data and the $T^{*}$ values inferred from and resistance data in figure \ref{figure11}. In addition these data suggest that magnetic field could be used to fine tune the value of $T^*$, if needed.

\section{Superconducting critical fields and anisotropy }

 \begin{figure}
 	\includegraphics[width=1.5\columnwidth]{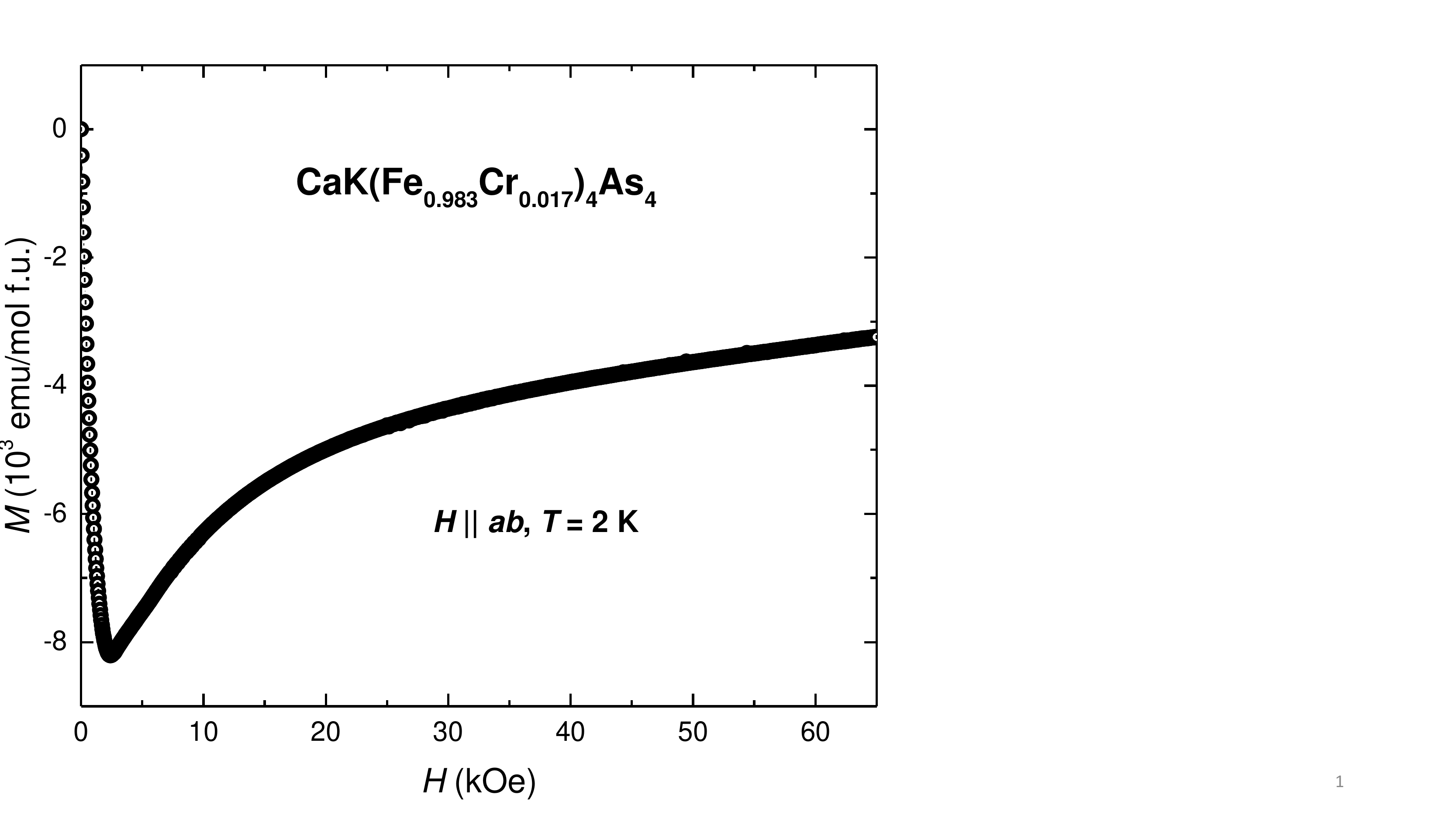}		
 	\caption{Magnetization of a single crystal of CaK(Fe$_{0.983}$Cr$_{0.017}$)$_{4}$As$_{4}$ as a function of magnetic field applied parallel to the crystallographic \textit{ab} plane at 2 K. In zero applied field cooling (ZFC) to 2 K so as to demagnetization is done at 60 K before cooling to minimize the remnant magnetic field.     \label{figure6}}
 \end{figure}

Superconductivity can be studied as a function of field (in addition to temperature and doping). Before we present our $H$$_{c2}$($T$) results, based on $R$($T,H$) data, it is useful to check the $M$($H$) data. We start with \textit{M}(\textit{H}) data for $x$ = 0.012, $T_c$ = 21.3 K, taken over a wide field range. The 2 K \textit{M}(\textit{H}) data for shown in figure \ref{figure6} is classically non-linear, showing a local minimum near $H$ $\sim$ 2.5 kOe. For $T$ = 2 K < $T_{c}$ the $H_{c2}$ value is clearly higher than the 65 kOe maximum field we applied (see discussion and figures below).~$H_{c1}$ can be inferred from the lower field \textit{M}(\textit{H}) data.

\begin{figure}
	\centering
	\begin{minipage}{0.44\textwidth}
		\centering
		\includegraphics[width=1.5\columnwidth]{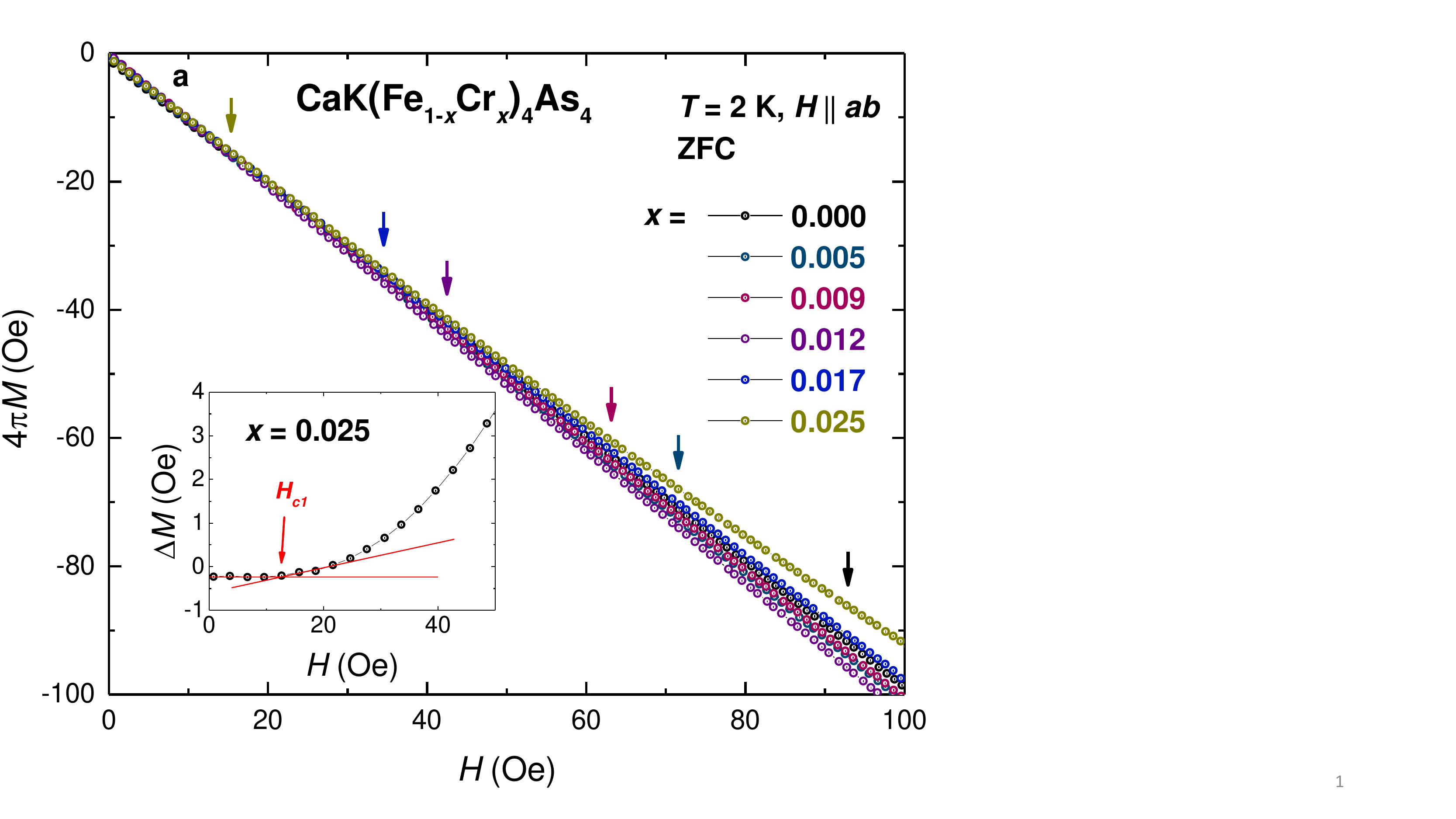}		
	\end{minipage}\hfill
	\centering
	\begin{minipage}{0.44\textwidth}
		\centering
		\includegraphics[width=1.5\columnwidth]{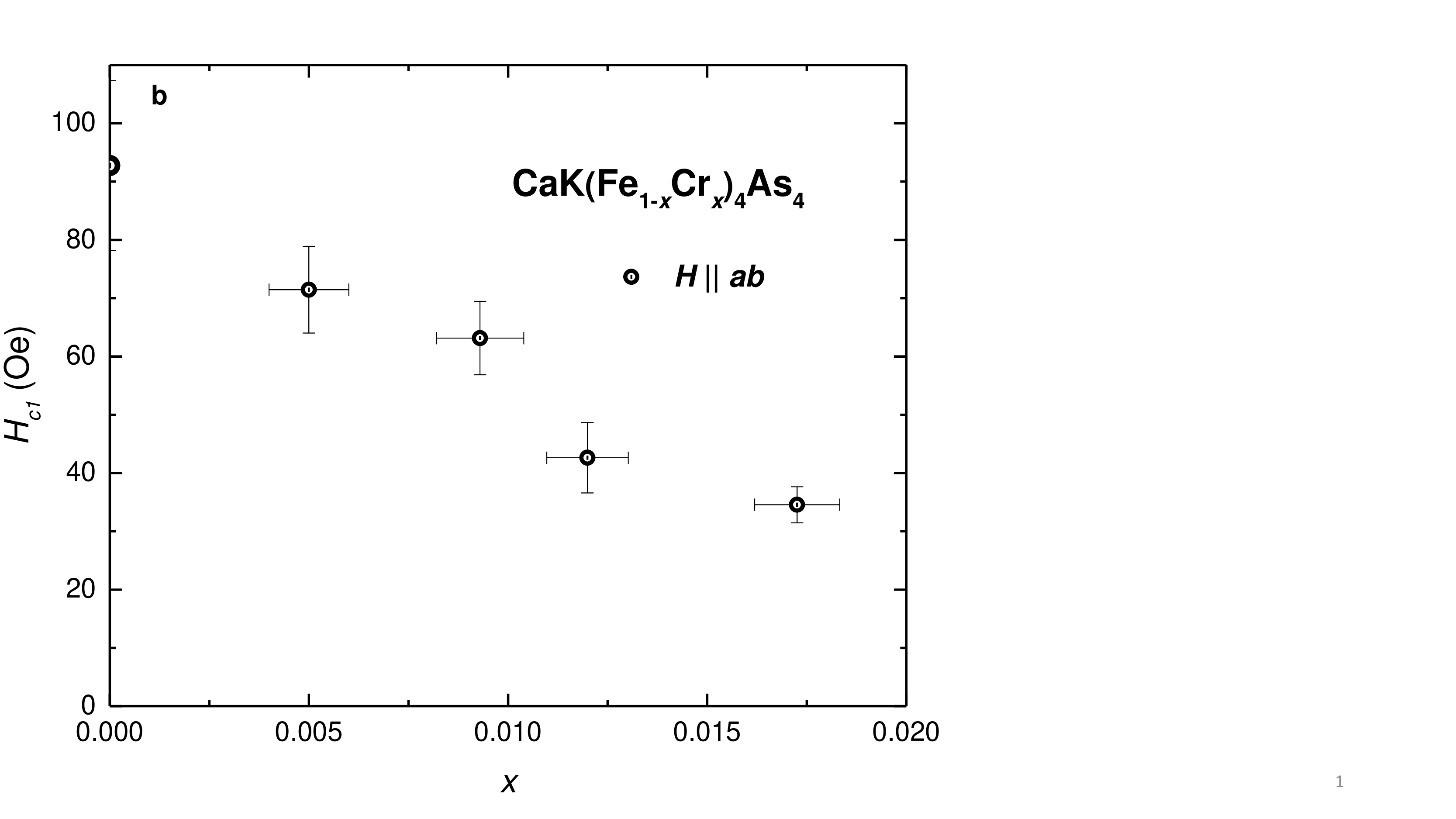}		
	\end{minipage}\hfill
	\caption{(a) Magnetization as a function of magnetic field applied parallel to the crystallographic \textit{ab} axis at 2 K for different substitution levels. Arrows mark the value of the magnetic field ($H_{c1}$) where $M$(\textit{H}) deviates from linear behavior. Inset shows the criterion we use to determine the $H_{c1}$ values. Remnant field of measurements are smaller than 1 Oe, which is consistent with $M$($H$) plots shown in figure. Temperature is ZFC to 2 K and demagnetization is done at 60 K before cooling to minimize the remnant magnetic field (b) $H_{c1}$ value versus $x$.}\label{HC12sum}
\end{figure}

In order to better estimate $H_{c1}$ values we performed low field, $M(H)$ sweeps at base temperature. In figure \ref{HC12sum}a we show the \textit{M}(\textit{H}) data for 0 $\leqslant$ $x$ $\leqslant$ 0.025 for $H$ $\leqslant$ 100 Oe. As $x$ increases the deviation from the fully shielded, linear behavior, that occurs at $H_{c1}$, appears at lower and lower fields. As shown in the inset of figure \ref{HC12sum}a, $\Delta$$M$ is determined by subtracting the linear, lowest field behavior of $4\pi M$ from $H$. Given the finite thickness of samples and field direction applied in $ab$ plane, there is a small demagnetizing factor ($N$ < 0.077), therefore, $H_{c1}$ is taken as the vortices start to enter the sample and determined as the point when the $M$($H$) data deviates from the linear, lowest field behavior. The non-zero value is due to remnant field of MPMS. The standard error of $H_{c1}$ comes from at least 4 different samples' measurements. Figure \ref{HC12sum}b shows $H_{c1}$ at 2 K with different substitution. $x$ = 0.025 is not shown in this plot since $T_c$ $\sim$ 7 K is close to 2 K. The data shown in figure \ref{HC12sum}b are roughly linear in $x$, but there may be a hint of a change in behavior near $x$ $\sim$ 0.01, where $T_c$ drops below $T^*$ in figure \ref{figure11}. This will be discussed further when examine the London penetration depth below.

\begin{figure}
	\centering
	\begin{minipage}{0.44\textwidth}
		\centering
		\includegraphics[width=2.7\columnwidth]{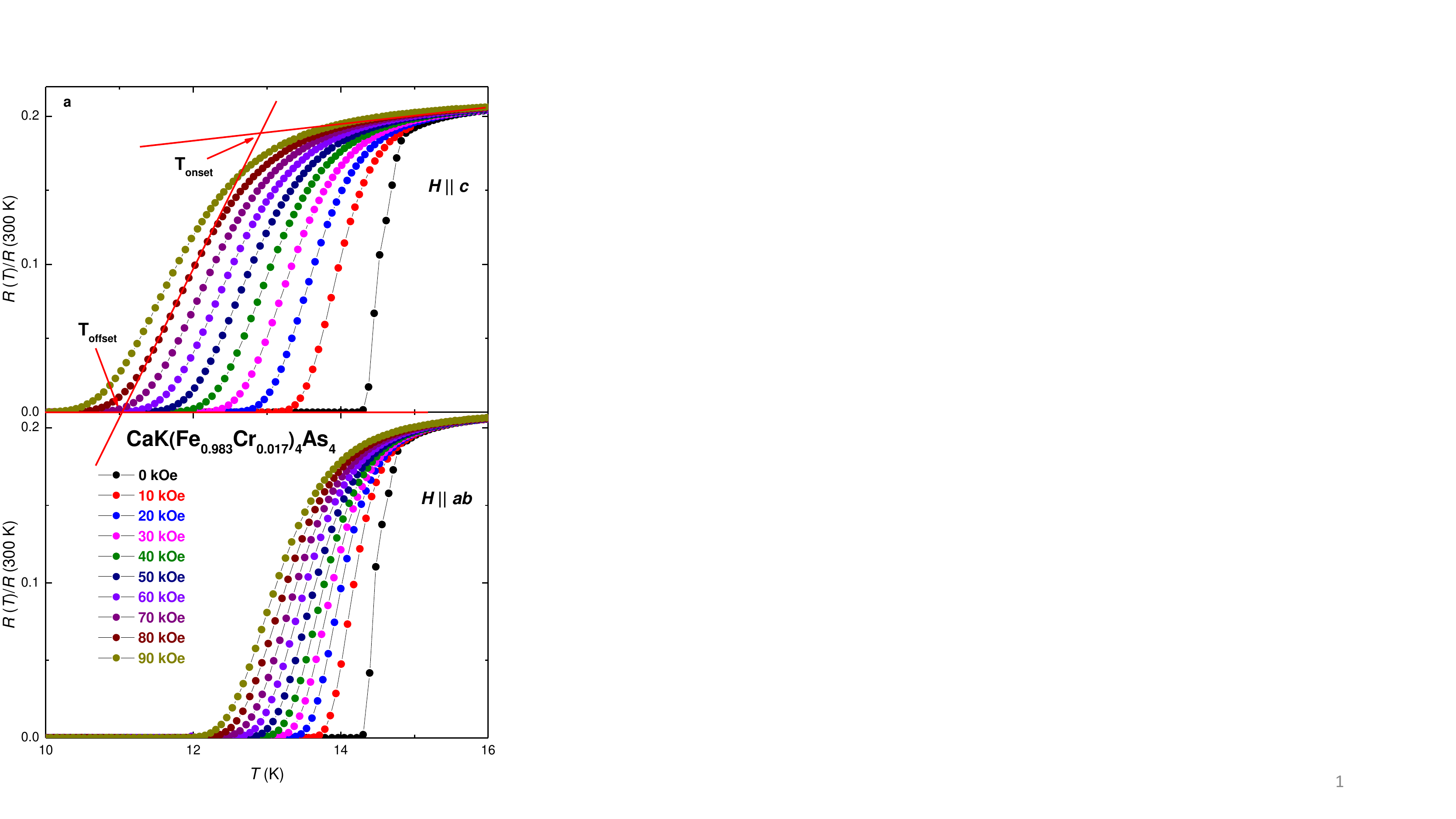}		
	\end{minipage}\hfill
	\centering
	\begin{minipage}{0.44\textwidth}
		\centering
		\includegraphics[width=1.5\columnwidth]{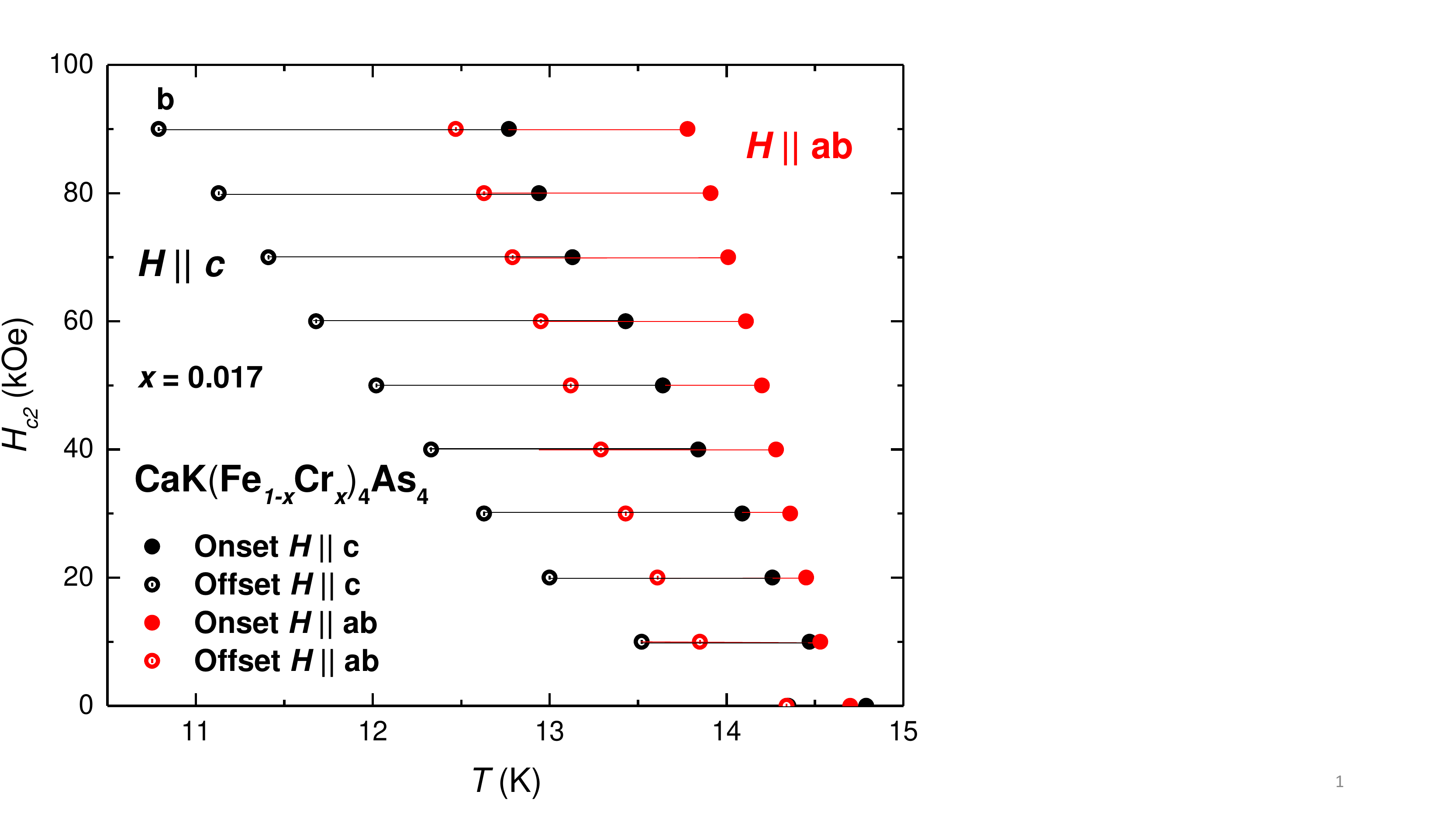}		
	\end{minipage}\hfill
	\caption{(a) Temperature-dependent electrical resistance of CaK(Fe$_{0.983}$Cr$_{0.017}$)$_{4}$As$_{4}$ single crystal for magnetic field parallel to the crystallographic \textit{c} axis (upper panel) and $ab$ plane (lower panel) for representative fields $\textit{H} \leq 90$ kOe. Onset and offset criteria for $T_c$ are shown by the red solid lines. (b) Anisotropic $H_{c2}(T)$ data determined for two single crystalline samples of CaK(Fe$_{0.983}$Cr$_{0.017}$)$_{4}$As$_{4}$ using onset criterion (solid) and offset criterion (hollow) inferred from the data shown in (a).}\label{figure7}
\end{figure}

In order to further study the effects of Cr substitution on the superconducting state, anisotropic $H_{c2}(T)$ data for temperatures near $T_{c}$ were determined for the substitution levels that have superconductivity. Figure \ref{figure7}a shows a representative set of $R$($T$) data taken for fixed applied magnetic fields, $H$ $\parallel$ $c$ axis and $ab$ plane $\leq$ 90 kOe for $x$ = 0.017. Figure \ref{figure7}a also shows an example of the onset and offset criteria used for the evaluation of $T_c$. Figure \ref{figure7}b presents the anisotropic $H_{c2}(T)$ curves for CaK(Fe$_{0.983}$Cr$_{0.017}$)$_{4}$As$_{4}$ single crystals with both $H$ || $c$ and $H$ || $ab$, showing both the onset (T$_{onset}$) and offset (T$_{offset}$) temperatures. For $x$ = 0.005, 0.009, 0.012 and 0.025, $H_{c2}(T)$ curves are shown in Appendix figures \ref{figure81} - \ref{figure84}. From $H_{c2}(T)$ plots, we can see that $T_{c}$ is only suppressed by about 4~K when 90~kOe magnetic field is applied, so the complete $H_{c2}(T)$ plots of the CaK(Fe$_{1-x}$Cr$_x$)$_4$As$_4$ compounds cannot be fully determined, however we still can observe several trends in these data.

\begin{figure}
 	\includegraphics[width=1.4\columnwidth]{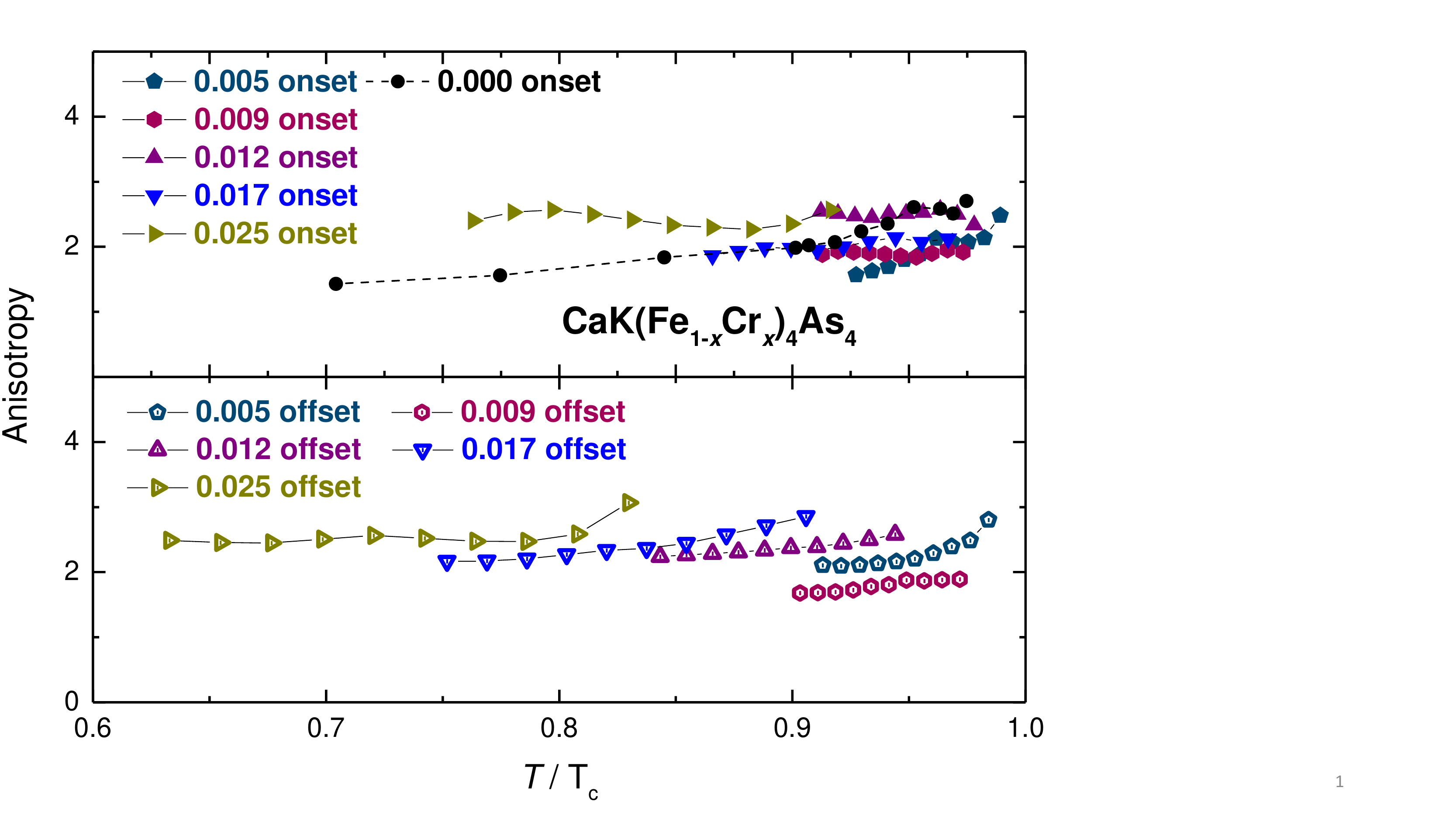}	
 	\caption{Anisotropy of the upper critical field, $\gamma = H_{c2}^{\parallel ab}(T)/H_{c2}^{\parallel c}(T)$, as a function of effective temperature, $T$/$T_{c}$, for CaK(Fe$_{1-x}$Cr$_{x}$)$_{4}$As$_{4}$ single crystals, using onset criterion (upper panel) and offset criterion (lower panel), inferred from the temperature-dependent electrical resistance data. The $T_{c}$ value used to calculate the effective temperature (\textit{T}/$T_{c}$) is the zero-field superconductivity transition temperature for each Cr-substitution levels (see figure \ref{figure11} above). Data CaKFe$_{4}$As$_{4}$ are shown in black circles with dashed line.  \label{figure12}}
 \end{figure}

Figure \ref{figure12} shows the temperature dependent anisotropy ratio, $\gamma = H_{c2}^{\parallel ab}(T)/H_{c2}^{\parallel c}(T)$, is around 2.5 for these samples over the 0.6 < $T$/$T_c$ < 1.0 range. This value is  similar to other 122 and 1144 materials {\color{blue}\cite{Ni20101,Meier2016,Mingyu01,Tanatar2010}}. $\gamma$ $\sim$ 2.5 is also qualitatively consistent with the estimated resistivity anisotropy ratio $\gamma_{\rho}=\rho_c/\rho_a\approx$$ 3- 6 $ at 300~K, increasing to $6- 10$ at $T$ = 0 K, with $\gamma_H \sim \sqrt{\gamma_{\rho}}$ {\color{blue}\cite{Murphy2013}}. Black circles and dashed line present the data for pure CaKFe$_{4}$As$_{4}$ {\color{blue}\cite{Meier2016}}. The anisotropies of coherence length and penetration depth are expected to be the same close to $T_c$, but can have opposite temperature dependence upon cooling below that {\color{blue}\cite{Kogan2019}}. However, almost no temperature dependence of $\gamma$ is seen in the temperature range measured. Based on that, the average values of gamma can be taken as a good estimate of both anisotropies at low temperature as well.

 

\begin{figure}
	\includegraphics[width=1.4\columnwidth]{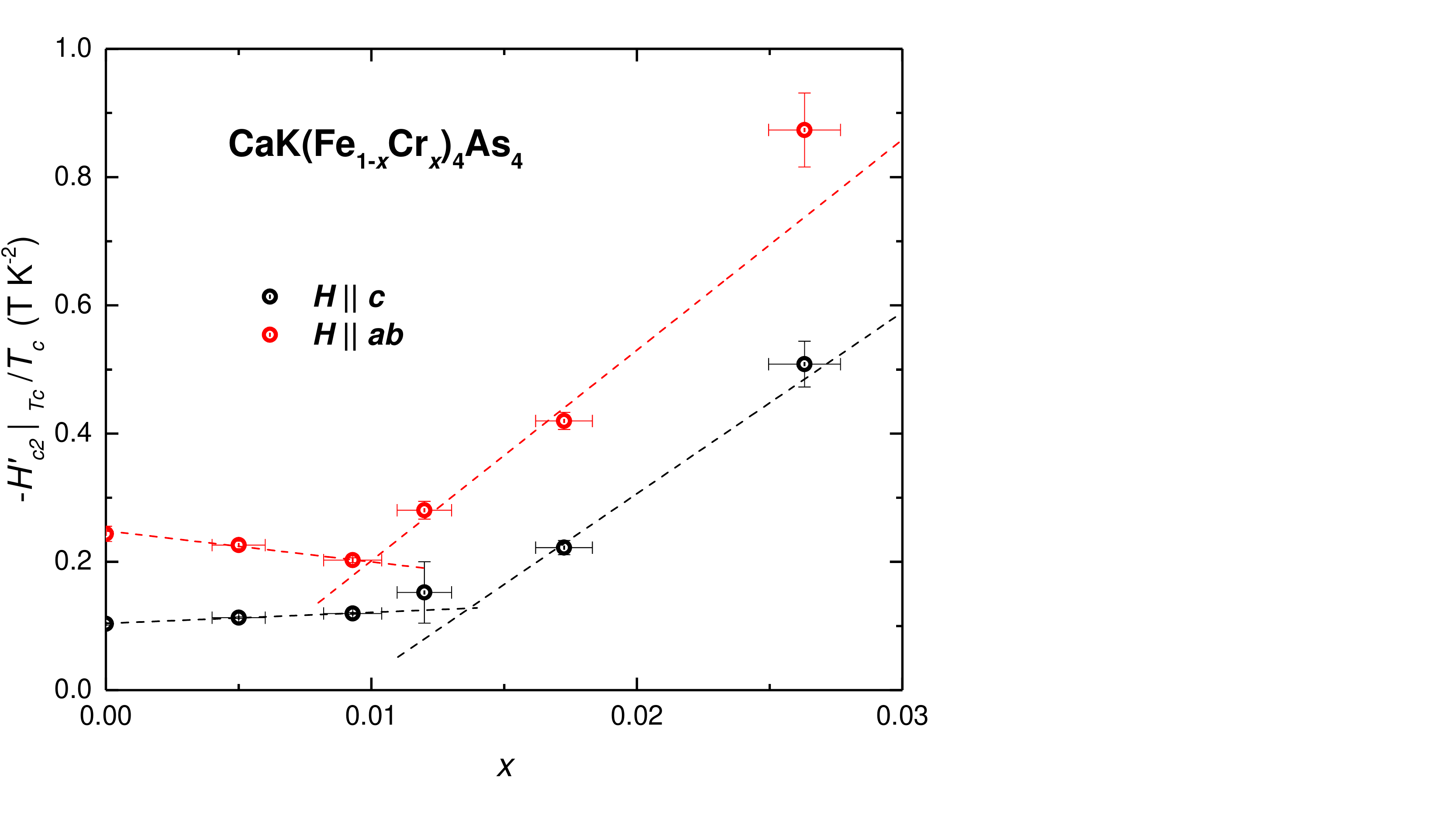}			
	\caption{Substitution dependence of slope of $H_{c2}$($T$) at $T_c$,~-$H^\prime_{c2}$($T$)/$T_c$, where $H^\prime_{c2}$($T$) is d$H_{c2}$($T$)/d$T$ at $T_c$. $T_c$ is determined by $T_{offset}$ from criteria shown in figure \ref{figure7}. Qualitatively similar, albeit somewhat weaker, results can be seen using $T_{onset}$ criterion data. Dash lines show the linear fit for $x$ < 0.01 and $x$ > 0.01, which indicate the change of trend of $H_{c2}$($T$) at $T_c$ versus $x$ as substitution increase. \label{figureHc2dt}}
\end{figure}

Given that we have determined $H_{c2}$($T$) for temperatures close to $T_c$, we can evaluate the $H^\prime_{c2}$($T$)/$T_c$ close to $T_{c}$, where $H^\prime_{c2}$($T$) is d$H_{c2}$($T$)/d$T$, specifically seeing how it changes as $T_c$ drops below $T^*$ with increasing x. Error of $H^\prime_{c2}$($T$)/$T_c$ comes from linear fit of  $H_{c2}$($T$) near the $T_c$. In the case of other Fe-based systems {\color{blue}\cite{Xiang2018,Kaluarachchi2016,Xiang2017,Taufour2014,Mingyu01}} clear changes in $H^\prime_{c2}$($T$)/$T_c$ were associated with changes in the magnetic sublattice coexisting with superconductivity (i.e. ordered or disordered). In figure \ref{figureHc2dt} we can see that there is a change in the x-dependence of $H^\prime_{c2}$($T$)/$T_c$ for $x$ $>$ 0.012, beyond which substitution level suppresses $T_c$ below $T^*$. Comparison with the slope change of $H_{c2}$ in the pressure-temperature phase diagram of CaK(Fe$_{1-x}$Ni$_{x}$)$_4$As$_4$ {\color{blue}\cite{Xiang2018}}, further suggest that this is probably related to changes in the Fermi surface, caused by the onset of the new periodicity associated with the AFM order.


\section{discussion and summary}

The $T$-$x$ phase diagram for CaK(Fe$_{1-x}$Cr$_x$)$_4$As$_4$ (Figure \ref{figure11}) is qualitatively similar to those found for Co-, Ni- and Mn- substituted CaKFe$_4$As$_4$, there is a clear suppression of $T_{c}$ with increasing Cr substitution as well as an onset of what is likely to be a AFM ordering for $x$ > 0.012. 

In figures \ref{HC12sum} and \ref{figureHc2dt} we presented measurements and analysis of $H_{c1}$ and $H_{c2}$ data. Whereas we see only subtle, if any effect of the onset of AFM ordering on $H_{c1}$ (figure \ref{HC12sum}b), there is a clear effect on $H_{c2}$ (figure \ref{figureHc2dt}). Using our $H_{c1}$ and $H_{c2}$ data we can extract information about the superconducting coherence length and London penetration depth as well.

\begin{figure}
	\centering
	\begin{minipage}{0.44\textwidth}
		\centering
		\includegraphics[width=1.5\columnwidth]{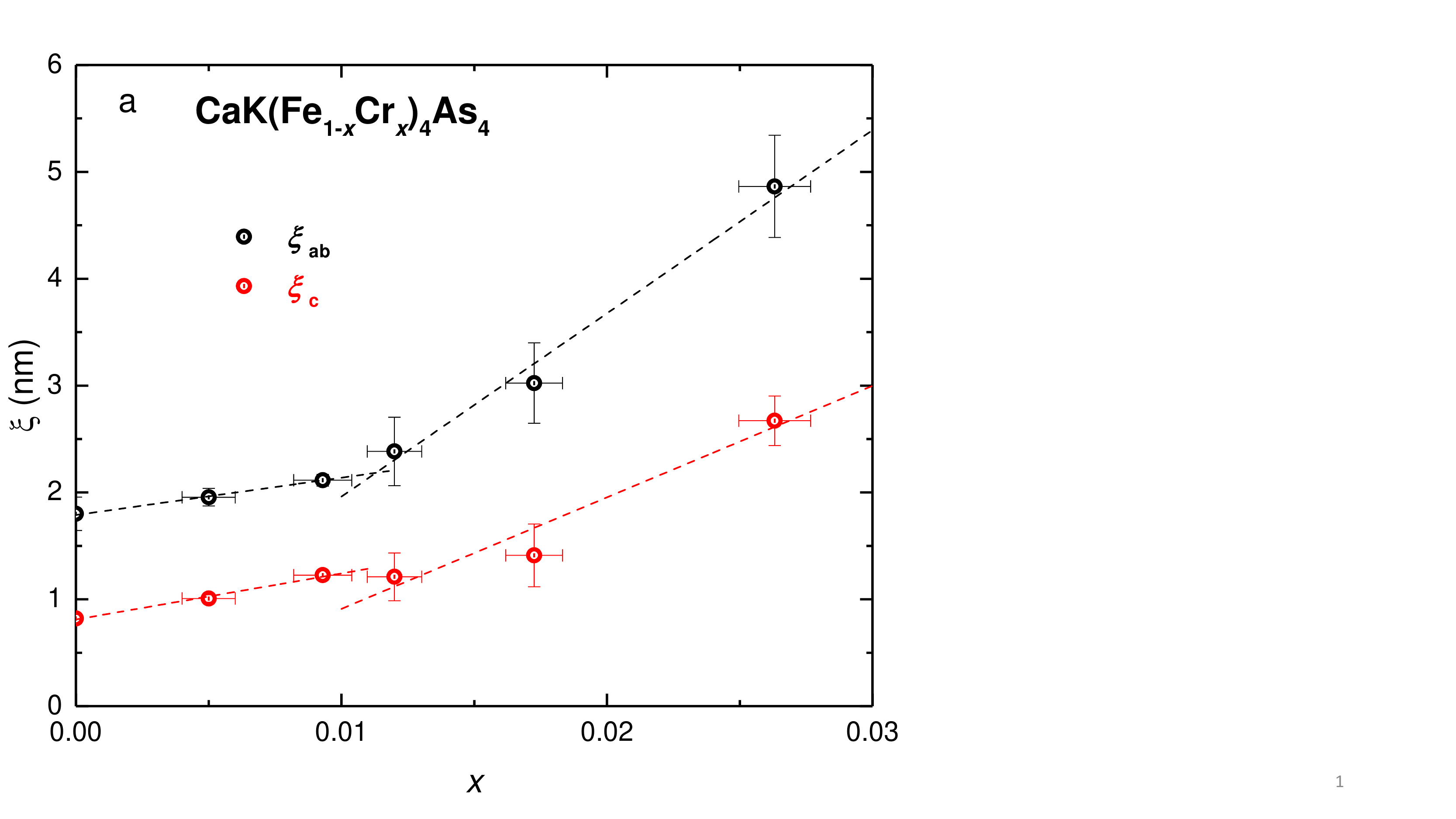}	
	\end{minipage}\hfill
	\begin{minipage}{0.44\textwidth}
		\centering
		\includegraphics[width=1.5\columnwidth]{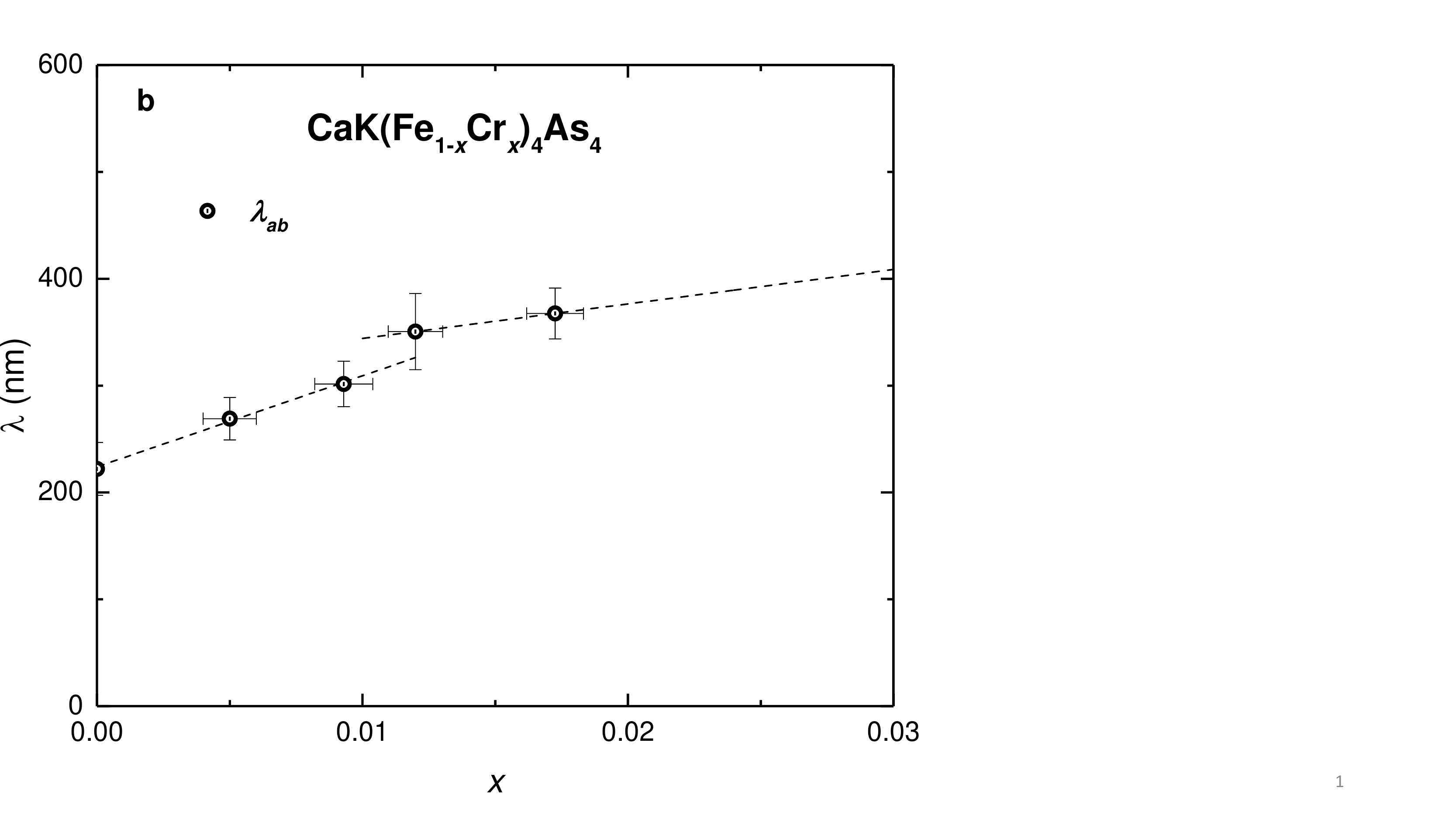}		
	\end{minipage}\hfill
	
	\caption{(a) Coherence length versus $x$ plot of CaK(Fe$_{1-x}$Cr$_x$)$_4$As$_4$ single crystals with applied field in $c$ direction and $ab$ plane. (b) shows the London penetration depth as a function of $x$ obtained by magnetic field applied parallel to the crystallographic \textit{ab} axis with different substitution levels. Dash lines show the linear fit for $x$ < 0.01 and $x$ > 0.01, which show the change of coherence lengths and London penetration depths at $T_c$ versus $x$ as substitution increase. \label{CoHL}}
\end{figure}

Figure~\ref{CoHL}a shows coherence length, $\xi$, of CaK(Fe$_{1-x}$Cr$_x$)$_4$As$_4$ as a function of $x$. $\xi$ is estimated by using the anisotropic scaling relations 0.69|d$H_{c2}^{||c}$/d$T$| = $\phi_0$/2$\pi$$\xi^2_{ab}$$T_c$ and 0.69|d$H_{c2}^{\perp c}$/d$T$| = $\phi_0$/2$\pi$$\xi_{c}\xi_{ab}$$T_c$ {\color{blue}\cite{Meier2016}}. We estimate $H_{c2}(0)$ with $H_{c2}(0)/(T_c|$d$H_{c2}(T_c)/$d$T|)$ = 0.69 {\color{blue}\cite{Kogan2013b}}. Figure \ref{CoHL}b shows the London penetration depth, $\lambda_{ab}$, as a function of $x$. Since, according to figure \ref{figure12}, $\gamma$ of $H_{c2}$ does not change much as the temperature decreases below T$_{c}$, the anisotropy of the penetration depth is estimated as the average of the anisotropy of $H_{c2}$ at low temperature. $\lambda_{ab}$ is obtained by using,
\begin{align}
		H_{c1}^{||ab} & =~\phi_0/(4\pi\lambda_{ab}\lambda_{c})(ln\kappa_{ab} + 0.5)\\
	& = \phi_0/(4\pi\lambda_{ab}^2\gamma)(ln(\lambda_{ab}/\xi_{ab}) + ln\gamma + 0.5)		
\end{align}
$\lambda_{c}/\lambda_{ab}$ = $\xi_{ab}/\xi_{c}$ = $H_{c2}^{\parallel ab}(T)/H_{c2}^{\parallel c}(T)$ = $\gamma$ = $1/\varepsilon$ with $\varepsilon$ being the angle-dependent anisotropy parameter and $\kappa_{ab}$ = $\sqrt{\lambda_{ab}\lambda_{c}/\xi_{ab}\xi_{c}}$ {\color{blue}\cite{Hu1972,Song2011,Blatter1994}}. Coherence lengths and penetration depths increase as substitution levels increase, and, given that $\xi$ depends on d$H_{c2}$/d$T$ and penetration depths depends on $H_{c1}$ and $\xi$. Both figure \ref{CoHL}a and b show breaks in behavior near $x$ $\sim$ 0.01, the substitution level at which $T^*$ emerges from below $T_c$.

\begin{figure}
	\centering
	\begin{minipage}{0.44\textwidth}
		\centering
		\includegraphics[width=1.5\columnwidth]{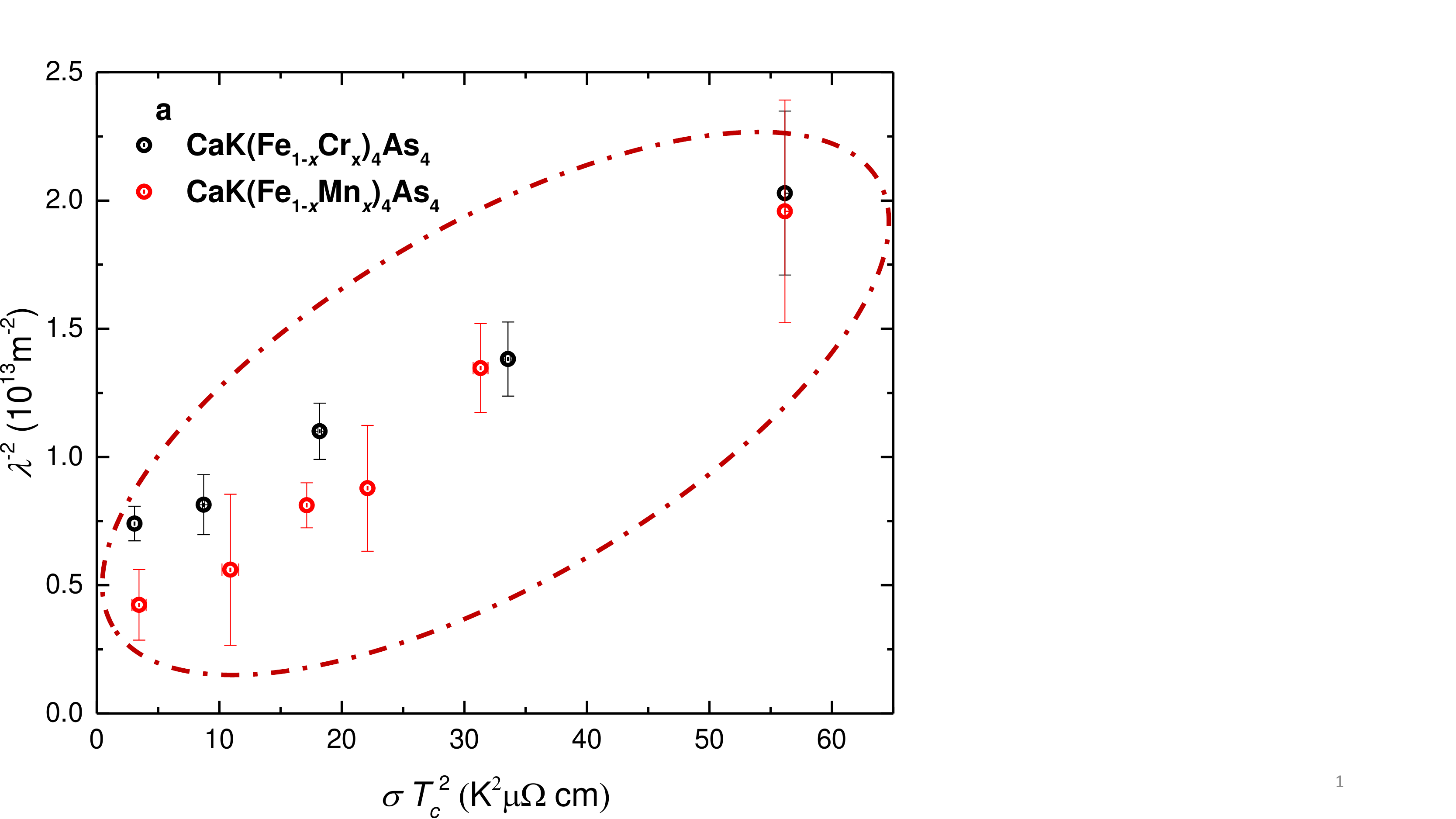}		
	\end{minipage}\hfill
	\begin{minipage}{0.44\textwidth}
	\centering
	\includegraphics[width=1.5\columnwidth]{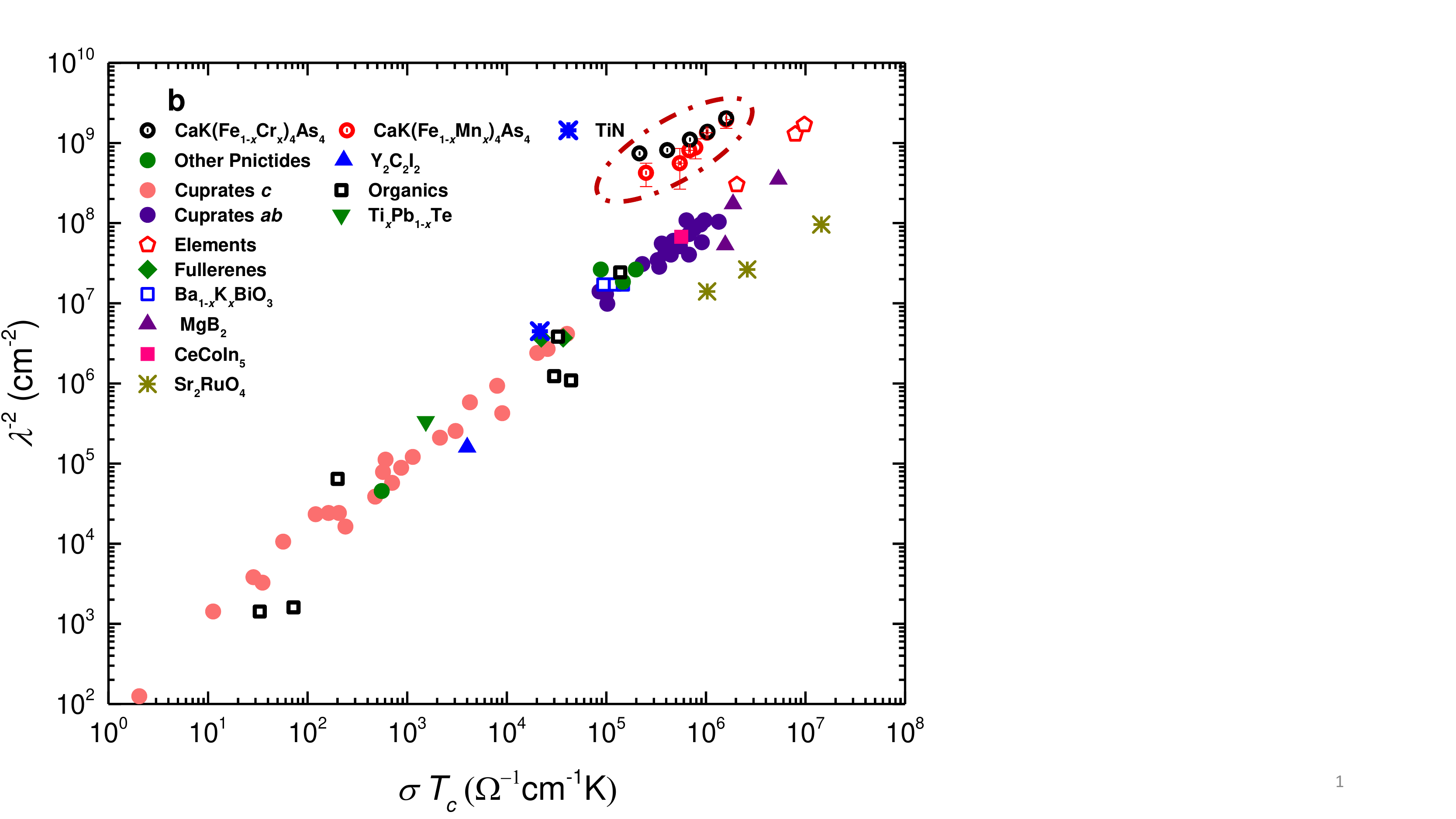}		
\end{minipage}\hfill
	\caption{(a) shows $\lambda ^{-2}$ versus $\sigma$ $T_c^2$ (Homes scaling) of Cr (black) and Mn (red) substitution of  CaKFe$_4$As$_4$. (b) shows Homes'scaling with other superconductors {\color{blue}\cite{Dordevic2013}}. Except CaK(Fe$_{1-x}$Cr$_x$)$_4$As$_4$ and CaK(Fe$_{1-x}$Mn$_x$)$_4$As$_4$, Homes' scaling is given by 1/$\lambda_s^2$ $\propto$ $T_c\sigma_{dc}$. $\sigma_{dc}$ is DC conductivity and data points are obtained from optical spectroscopes. Red dash-dot ellipse marks the Mn and Cr substituted 1144 data points. \label{Pena}}
\end{figure}

Figure \ref{Pena}a shows $\lambda^{-2}$ versus $\sigma$ $T_c^2$ in Cr and Mn substitution of CaKFe$_4$As$_4$, where $\sigma$ is normal state conductivity which was measured just before $T_c$ (Resistivity plot is shown in appendix figure \ref{Resistivity}). Whereas the CaK(Fe$_{1-x}$Mn$_x$)$_4$As$_4$ data roughly follow the behavior associated with the Homes type scaling in the presence of pair breaking {\color{blue}\cite{Dordevic2013,Kogan2013a,Kogan2013}}. The CaK(Fe$_{1-x}$Cr$_x$)$_4$As$_4$ data also follows the Homes scaling with slightly different slope. Figure \ref{Pena}b shows the Homes' scaling of superconductors on a log-log scale {\color{blue}\cite{Dordevic2013}}. Other pnictides are: Ba(Fe$_{0.92}$Co$_{0.08}$)$_4$As$_4$ and Ba(Fe$_{0.95}$Ni$_{0.05}$)$_4$As$_4$. Cuprates are YBa$_2$Cu$_3$O$_{6+y}$. On this log-log scale, both the CaK(Fe$_{1-x}$Cr$_x$)$_4$As$_4$ and CaK(Fe$_{1-x}$Cr$_x$)$_4$As$_4$ data sets agree rather well with other data, although being shifted up somewhat. It should be noted that the other data were determined from optical measurements {\color{blue}\cite{Dordevic2013}} and differences in criteria as well as measurement techniques may be responsible for the offset.

The $T$-$x$ phase diagrams of $T$ = Co and Ni substitutions in CaK(Fe$_{1-x}$$T$$_{x}$)$_{4}$As$_{4}$ scaled almost exactly as a function of band filling change (i.e. when each Ni atom brings two extra electrons and each Co atom brings only one extra electron) {\color{blue}\cite{Meier2018,Meier2019T}}. This led to the conclusion that for electron doping of CaKFe$_{4}$As$_{4}$ the number of electrons added was the control parameter for both the stabilization of magnetic ordering as well as for the suppression of superconductivity. This scaling did not seem to work for the case of Mn substitution {\color{blue}\cite{Mingyu01}}, but with only one "hole-like" transition metal substitution, it was hard to make clear conclusions.

\begin{figure}
	\centering
\begin{minipage}{0.44\textwidth}
\includegraphics[width=1.5\columnwidth]{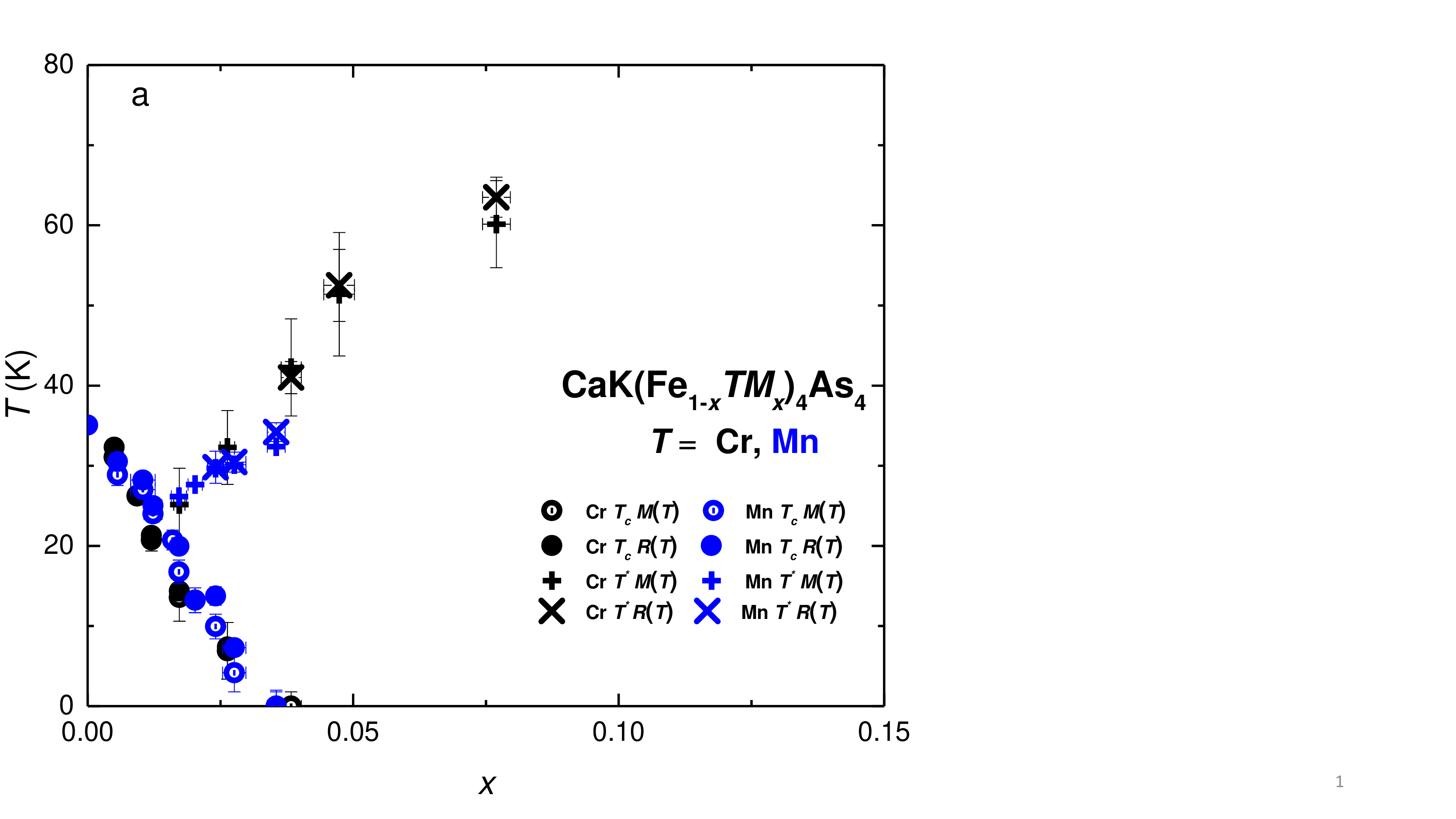}
\end{minipage}\hfill
\centering
\begin{minipage}{0.44\textwidth}
	\includegraphics[width=1.5\columnwidth]{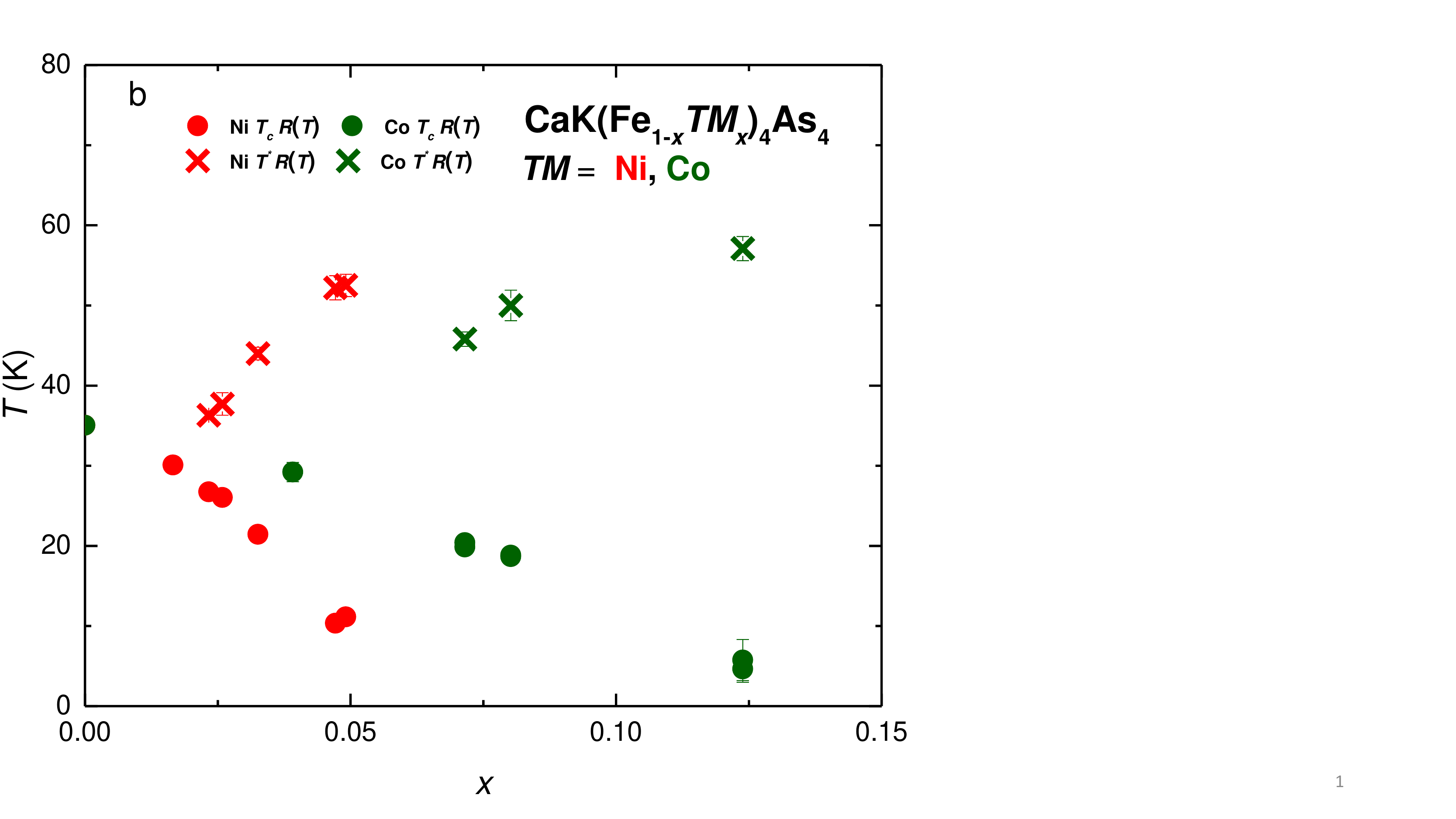}
\end{minipage}\hfill
\centering
\begin{minipage}{0.44\textwidth}
	\centering
	\includegraphics[width=1.5\columnwidth]{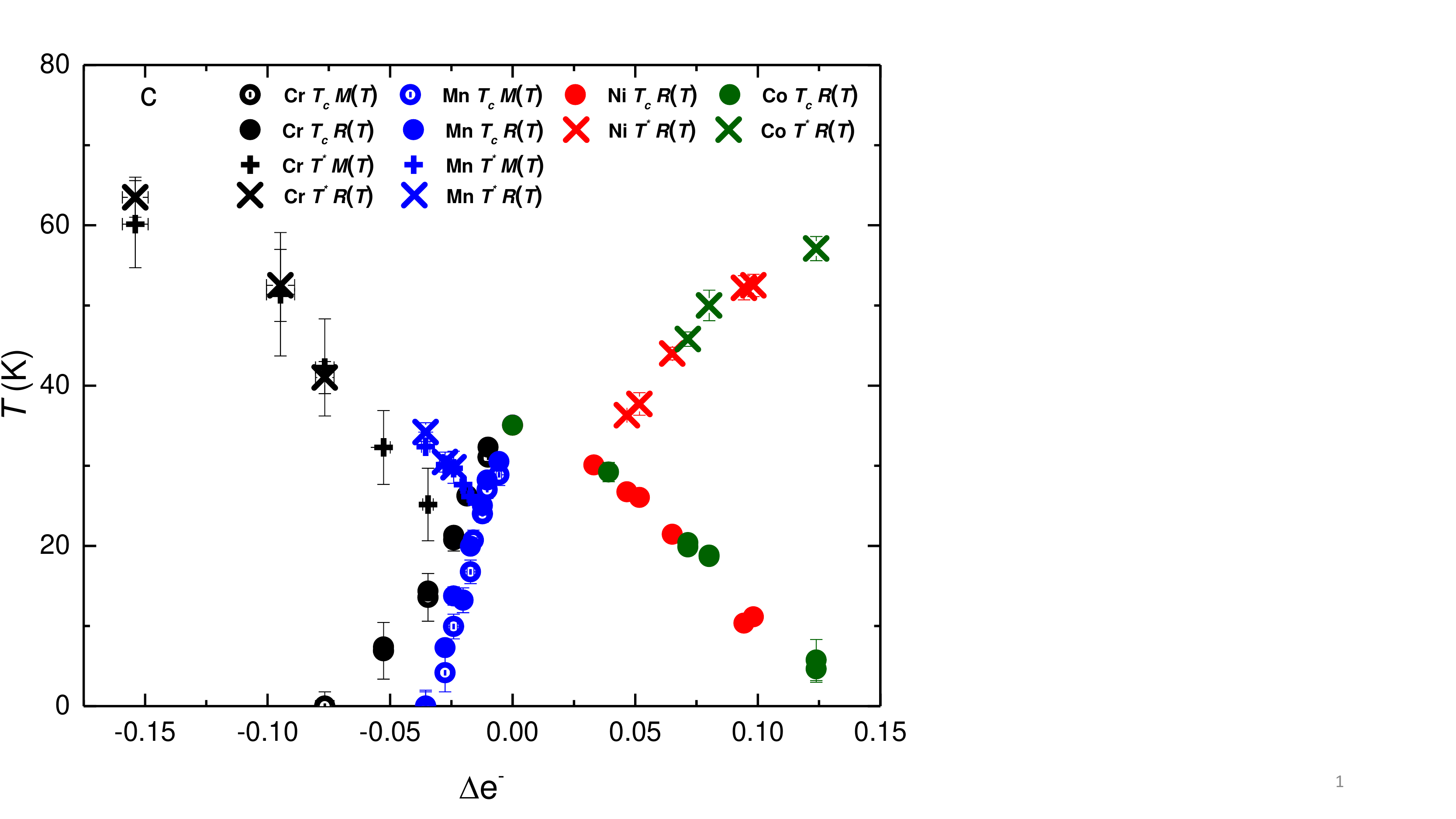}		
\end{minipage}\hfill
\caption{(a)(b) Temperature versus substitution level x phase diagram of CaK(Fe$_{1-x}T_{x}$)$_{4}$As$_{4}$ single crystals, $T$ = Cr, Mn, Ni and Co {\color{blue}\cite{Meier2018}}. (c) Temperature versus $\Delta e^-$ change of electrons phase diagram of CaK(Fe$_{1-x}T_{x}$)$_{4}$As$_{4}$ single crystals, $T$ = Cr, Mn, Ni and Co The circular symbols denote the $T_{c}$ phase transitions and the cross-like symbols denote the $T^{*}$ and $T_N$ phase transitions, which are obtained from resistance and magnetic moment measurements, denoted as "$R$($T$)" and "$M$($T$)".\label{figure31}}
\end{figure}

When Cr is substituted into CaKFe$_{4}$As$_{4}$, though, there is a qualitatively similar suppression of superconductivity as well as the stabilization of magnetic order, as was found for $T$ = Mn, Co and Ni. However there is a clear and important difference on a quantitative level, as shown in figure \ref{figure31}a. The $T$-$x$ phase diagram of CaK(Fe$_{1-x}$Cr$_x$)$_4$As$_4$ is essentially identical to that of CaK(Fe$_{1-x}$Mn$_x$)$_4$As$_4$. This is very different behavior from, figure \ref{figure31}b, that seen for electron doped 1144 where electron count seemed to be the key variable.

In figure \ref{figure31}c the CaK(Fe$_{1-x}T_{x}$)$_{4}$As$_{4}$ phase diagrams for $T$ = Cr, Mn, Co and Ni are plotted on the same $T$ and $\Delta$e$^{-}$ axes. Comparison of figures \ref{figure31}a, b and c reveal a clear and striking difference between the hole and electron doped  CaK(Fe$_{1-x}T_{x}$)$_{4}$As$_{4}$ systems. Whereas for the electron doped ($T$ = Co, Ni) there is very clear scaling with number of added electrons, for hole doped ($T$ = Mn, Cr) there is clear scaling with the number of substituted atoms, $x$. These striking differences in the phase diagrams beg the question of what is different between the two types of substitution.  When there was only data on Mn substitution to compare with the Co and Ni substitutions, one possible explanation could be based on an asymmetric density of electronic states on either side of $E$$_{F}$.  Given that the Cr- and Mn- substituted phase diagrams scale with $x$ rather than $e^-$, this is no longer a possibility.

A different approach to these data is to note that there is one other clear difference between the Mn and Cr substitutions as compared to the Co and Ni ones.  Mn and Cr clearly bring local-moment-like behavior as manifest by their conspicuous, high temperature, Curie tails that grow with increasing $x$.  This is absent for the Co and Ni substitutions. The effective moments coming from the Mn and Cr Curie tails ($\sim$ 5 $\mu_{B}$ {\color{blue}\cite{Mingyu01}} and $\sim$ 4 $\mu_{B}$ respectively) are consistent with Mn$^{3+}$ and Cr$^{3+}$ valencies. The Cr and Mn appear to behave like local moment impurities. As such it is not surprising that they lead to a stronger suppression of $T_c$ (via Abrikosov - Gor'kov pair breaking {\color{blue}\cite{A.A.AbrikosovandL.P.Gorkov1961}}). In a similar manner, it is not surprising that the addition of relatively large, local moment impurities to an itinerant, relatively small moment system helps to stabilize magnetic order. Given that the size of the local moments are similar, it is not surprising that we find that the $T$-$x$ phase diagrams scale well. The fact that CaKFe$_{4}$As$_{4}$ manifests rather bi-modal responses to $T$ substitution for $T$ = Co and Ni versus $T$ = Mn and Cr is consistent with the growing understanding of many of the Fe-based superconductors{\color{blue}\cite{Ni20101,Thaler2012}} families as manifesting properties in between those of a wide band metal (which would support rigid band shifting) and a more ionic- or Zintl-like compound (that would support valence-counting-like behavior).

In summary, we have been able to grow and study the CaK(Fe$_{1-x}$Cr$_x$)$_4$As$_4$ system. Based on magnetic and transport measurements, we assemble a $T$-$x$ phase diagram that clearly shows the suppression of the superconducting $T_{c}$ with the addition of Cr, with $T_{c}$ dropping from 35 K for $x$ = 0  to zero for $x$ $\sim$ 0.03, as well as the stabilization of magnetic order for $x$ > 0.012, with 22 K $\leq$ $T^{*}$ $\leq$ 60 K. As $x$ becomes greater than 0.012 and $T_{c}$ becomes less than $T^{*}$, a clear change in the behavior of $H^\prime_{c2}$($T$)/$T_c$ and the associated superconducting coherence length, $\xi$, can be seen. These are associated with the probable changes in the Fermi surface that accompany the AFM ordering at $T^{*}$. Comparable features in $H_{c1}$ or the London penetration depth are not clearly resolvable.

The $T$-$x$ phase diagram for CaK(Fe$_{1-x}$Cr$_x$)$_4$As$_4$ is qualitatively identical to CaK(Fe$_{1-x}$Mn$_x$)$_4$As$_4$ phase diagram due to the effect of local moment of Cr and Mn. Similarity of phase diagrams also indicate importance of the local moment impurity on hole doped 1144, which is different from the electron doped 1144. On the other hand, both hole and electron doping in 1144 stabilize antiferromagnetic ordering with increasing substitution level and the suppression of $T_{c}$ in CaK(Fe$_{1-x}$Cr$_x$)$_4$As$_4$ is faster than Ni- and Co-1144.

\begin{acknowledgements}
	We thank B. Kuthanazhi for useful discussions. Work at the Ames National Laboratory was supported by the U.S. Department of Energy, Office of Science, Basic Energy Sciences, Materials Sciences and Engineering Division. The Ames National Laboratory is operated for the U.S. Department of Energy by Iowa State University under Contract No. DE-AC02-07CH11358.
\end{acknowledgements}
\renewcommand\refname{[References]}
\bibliographystyle{apsrev}
\bibliography{Cr0209}

\clearpage
\section{Appendix}
\begin{figure}[H]
	\centering
	\begin{minipage}{0.44\textwidth}
		\centering
		\includegraphics[width=1.5\columnwidth]{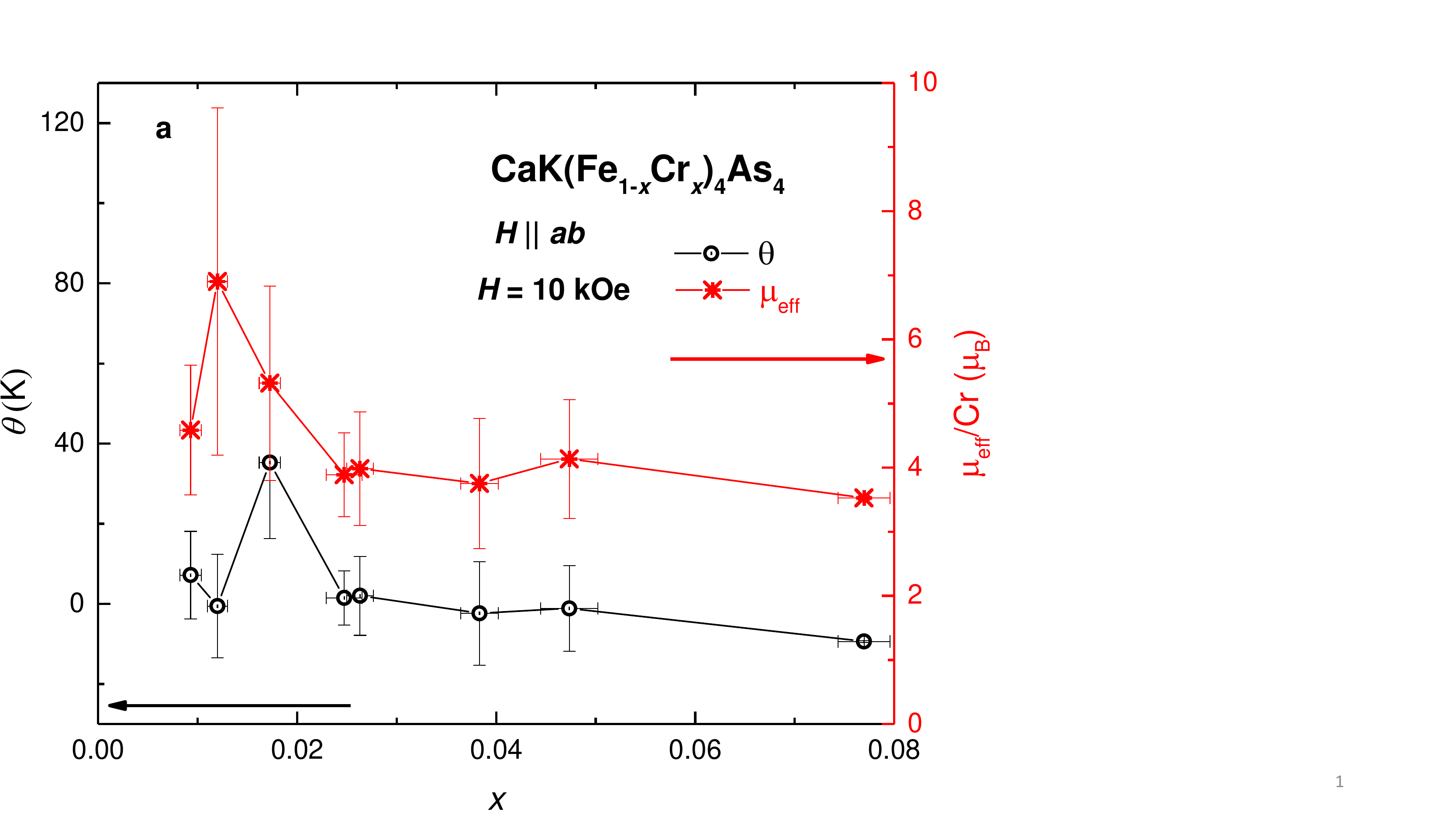}		
	\end{minipage}\hfill
	\centering
	\begin{minipage}{0.44\textwidth}
		\centering
		\includegraphics[width=1.5\columnwidth]{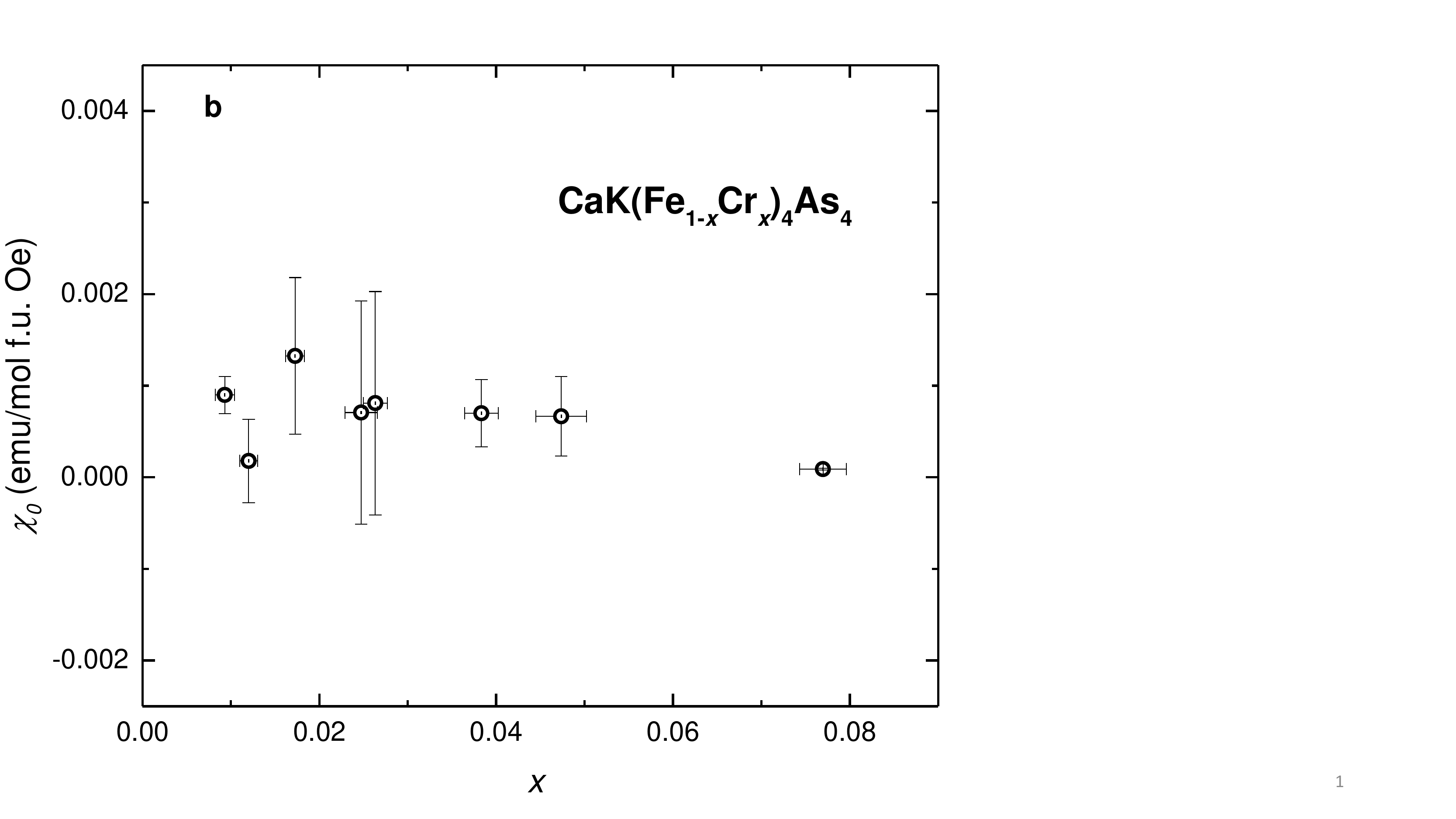}		
	\end{minipage}\hfill
	\caption{(a) shows Curie-Weiss temperature, $\theta$, effective moment, $\mu_{eff}$, obtained from Curie-Weiss fit to the difference magnetization ($\Delta M^{\prime}$) between CaKFe$_{4}$As$_{4}$ and CaK(Fe$_{1-x}$Cr$_{x}$)$_{4}$As$_{4}$ single crystals as a function of temperature for with a field of 10 kOe applied parallel to the crystallographic \textit{ab} plane. (b) shows temperature-independent susceptibility , $\chi_0$.  \label{figure 13}}
\end{figure}

The magnetization plots shown in figure \ref{figure3} have the appearance of Curie-Weiss tails potentially associated with the Cr-substitution. We fit the magnetization difference ($\Delta M^{\prime}$) between CaK(Fe$_{1-x}$Cr$_{x}$)$_{4}$As$_{4}$ and CaKFe$_{4}$As$_{4}$ single crystals as a function of temperature from 20~K above $Tc$ to 250~K with a field of 10~kOe applied parallel to the crystallographic \textit{ab} plane by a C/($T$+$\theta$) + $\chi_{0}$ function assuming that tail behavior is only due to Cr. Figure \ref{figure 13} shows the result of fitting. The value of $\mu_{eff}$ is around 4 $\mu_{B}$. The fitting results comes from at least 4 different samples $M$($T$) measurements. The values of $\theta$, $\mu_{eff}$ and $\chi_0$ become more stable after $x$ = 0.17. Instability for small substitution levels may come from difficulty of fitting small tail-like contribution and when $x$ = 0.005, there is no clear tail shown.

\begin{figure}[H]
	\includegraphics[width=1.5\columnwidth]{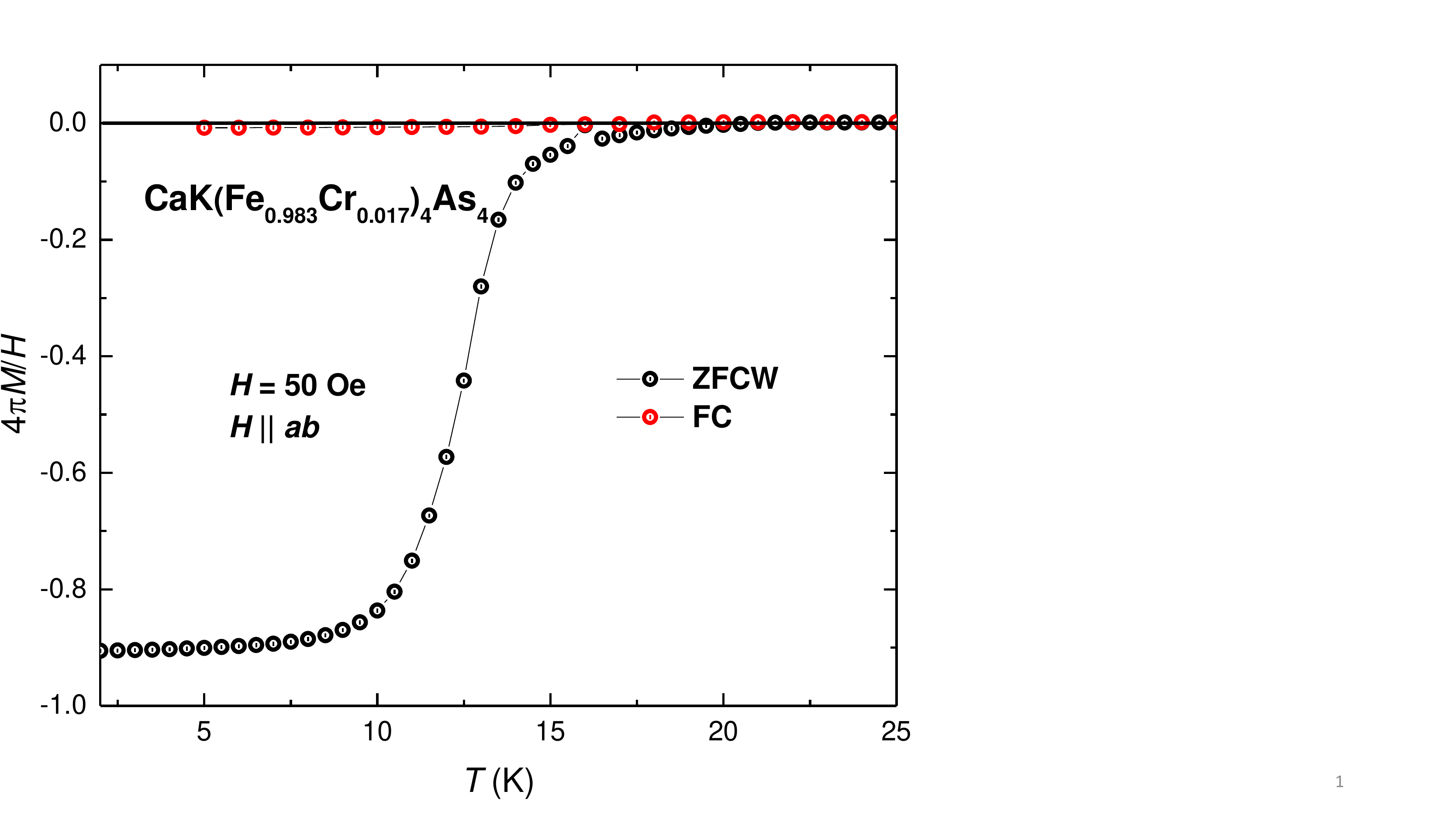}
	\caption{Zero-field-cooled-warming (ZFCW) and Field-cooled (FC) low temperature magnetization as a function of temperature for CaK(Fe$_{0.983}$Cr$_{0.017}$)$_{4}$As$_{4}$ single crystals with a field of 50~Oe applied parallel to the crystallographic \textit{ab} plane. $M$ is the volumetric magnetic moment with cgs unit emu cm$^{-3}$ or Oe.}  \label{50OeFCZFCz}
\end{figure}

Figure \ref{50OeFCZFCz} shows Zero-field-cooled-warming (ZFCW) and Field-cooled-warming (FC) low temperature magnetization as a function of temperature for CaK(Fe$_{0.983}$Cr$_{0.017}$)$_{4}$As$_{4}$ single crystals with a field of 50 Oe applied parallel to \textit{ab} plane. The large difference between ZFCW and FCW is consistent with the large pinning found even in pure CaKFe$_4$As$_4$\cite{Tomioka1993}.

\begin{figure}[H]
	\includegraphics[width=1.5\columnwidth]{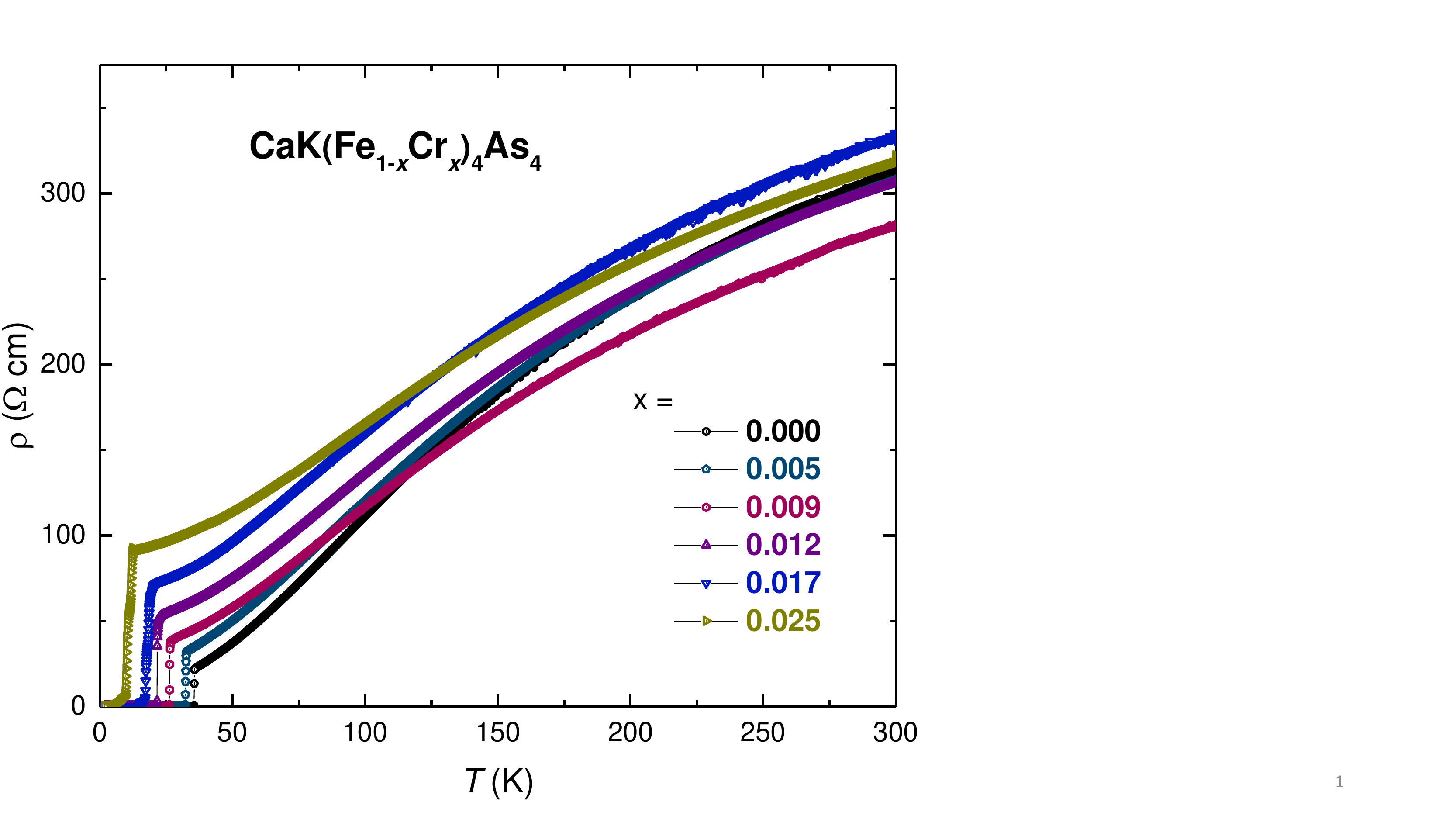}
	\caption{Temperature dependence of Resistivity, $\rho$, of CaK(Fe$_{1-x}$Cr$_{x}$)$_{4}$As$_{4}$ single crystals showing the suppression of the superconducting transition $T_c$.}  \label{Resistivity}
\end{figure}

Figure \ref{Resistivity} shows temperature dependence of resistivity, $\rho$, of CaK(Fe$_{1-x}$Cr$_{x}$)$_{4}$As$_{4}$ single crystals with $x$ < 0.38. Thickness is estimated by the density of pure CaKFe$_{4}$As$_{4}$, the mass and area of plate-like samples. The superconducting transition temperature is suppressed and resistivity before $T_c$ is increased by increasing substitution. Given the inevitably large geometric errors associated with the precised determination of length between voltage contacts as well as the sample thickness we consider the uncertainty in our resistivity values to be on the order of 20$\%$ and, as such we plot the data as R(T)/R(300 K) in the main text.

\begin{figure}[H]
	\centering
	\begin{minipage}{0.44\textwidth}
		\centering
		\includegraphics[width=1.5\columnwidth]{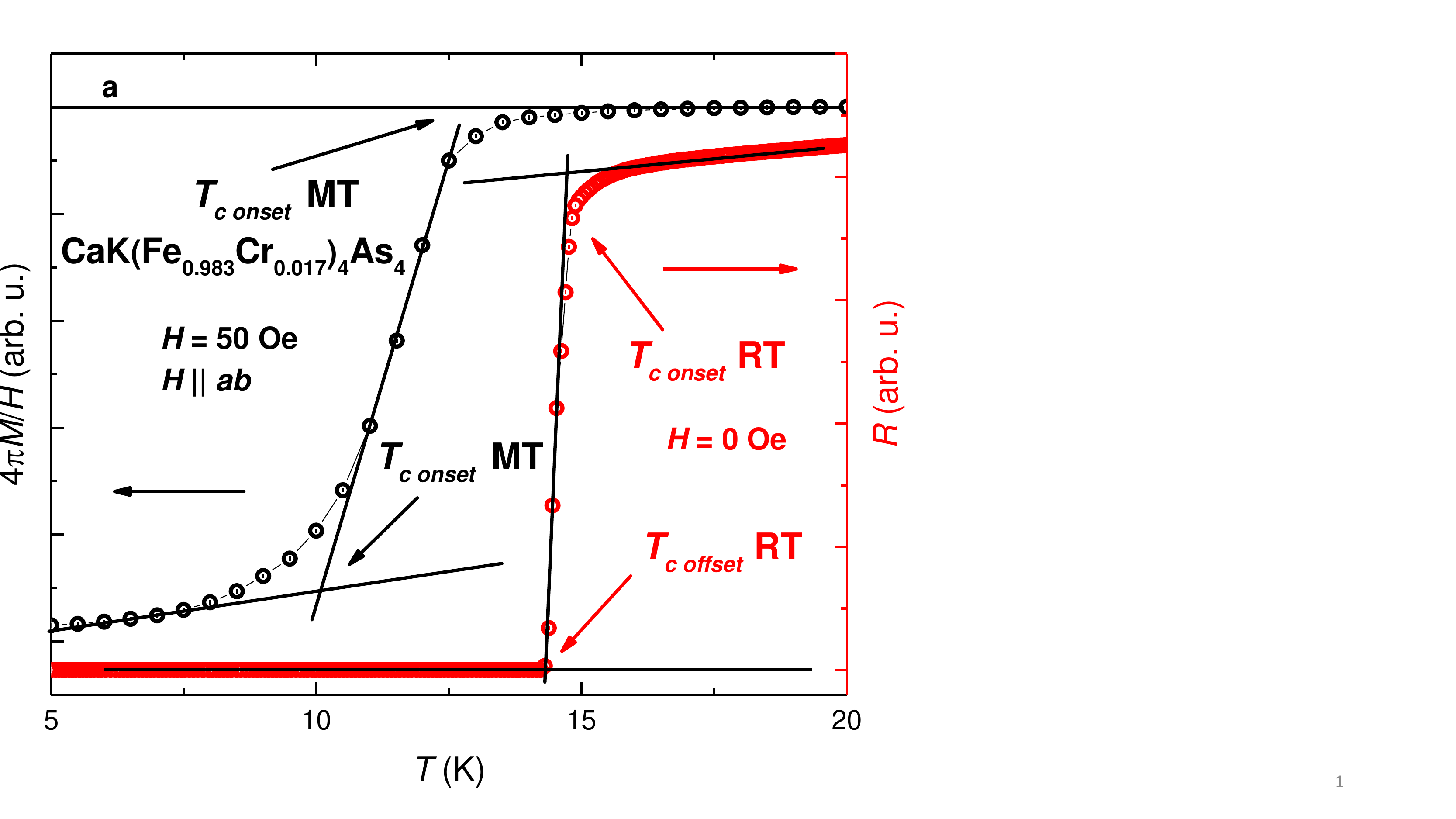}		
	\end{minipage}\hfill
	\begin{minipage}{0.42\textwidth}
		\centering
		\includegraphics[width=1.5\columnwidth]{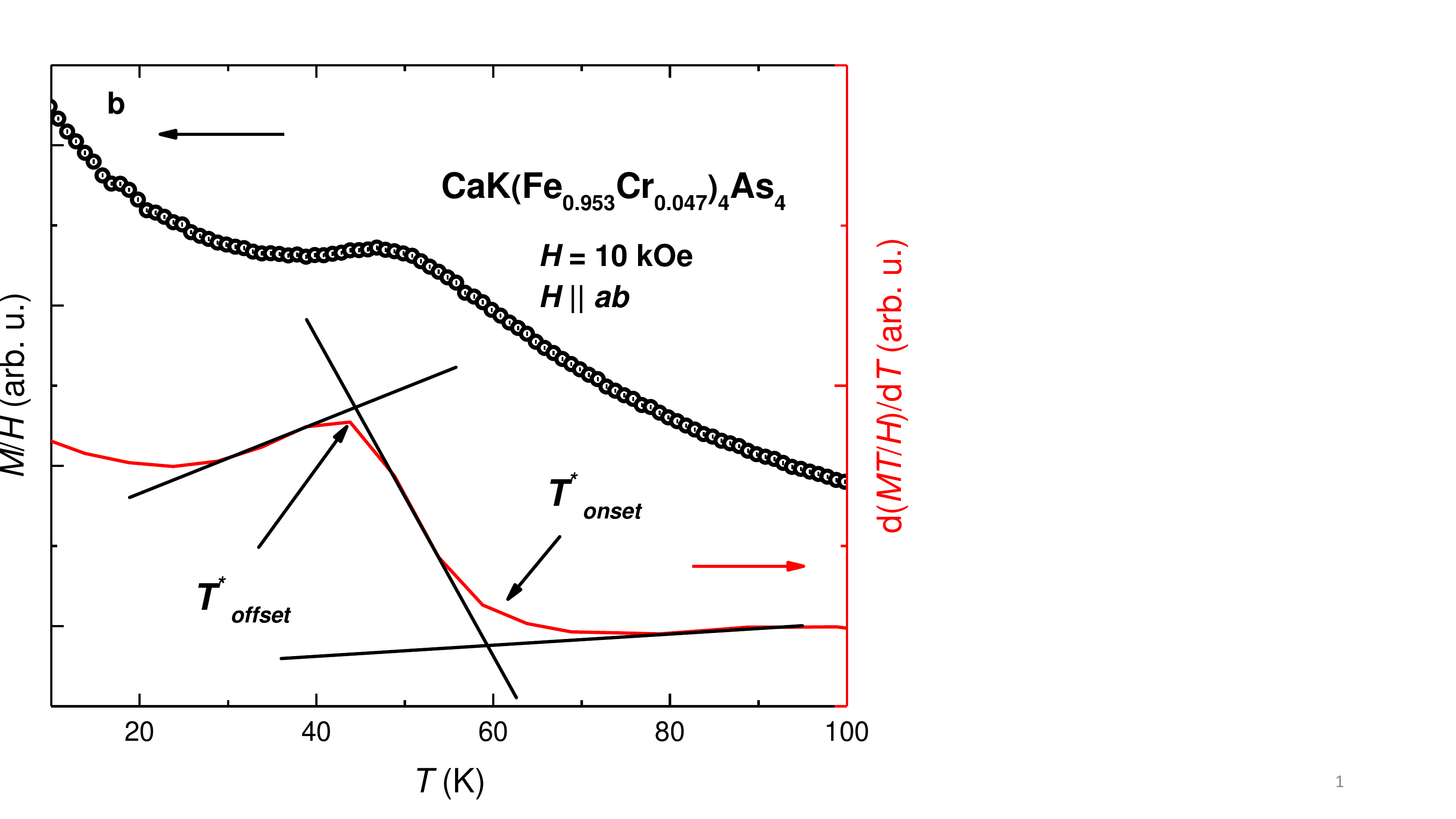}
		
	\end{minipage}\hfill
	\begin{minipage}{0.44\textwidth}
		\centering
		\includegraphics[width=1.5\columnwidth]{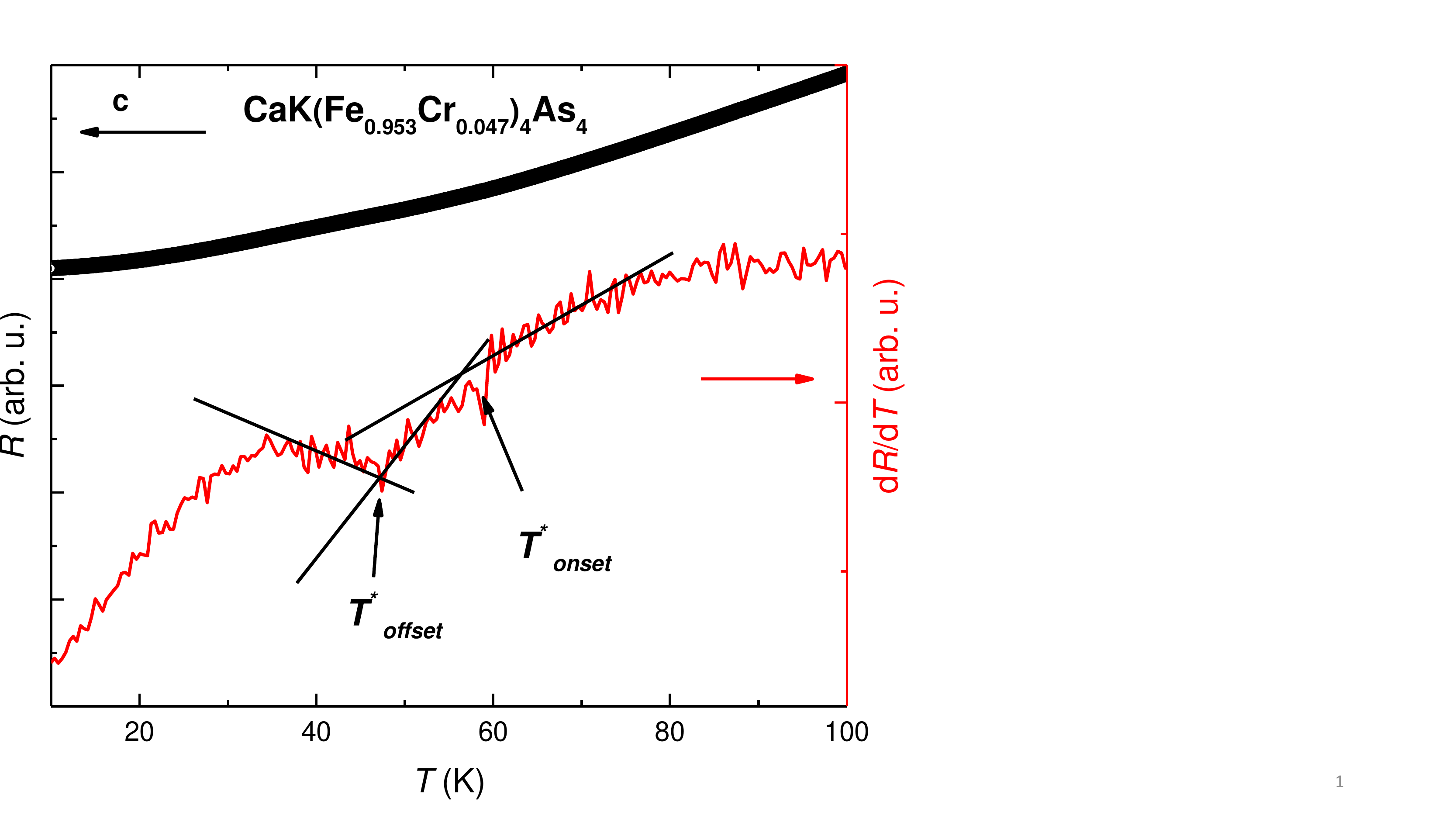}
		\caption{Onset and offset criteria for $T_c$ and $T^*$ based on magnetization and resistance measurement.   \label{figure92}}
	\end{minipage}
\end{figure}

Criteria for inferring $T_{c}$ and $T^{*}$ are shown in figure \ref{figure92}. 
 For $T_{c}$ (figure \ref{figure92}a) we use an onset criterion for \textit{M}(\textit{T}) data and an offset criterion for \textit{R}(\textit{T}) data. As is often the case, these criteria agree well, especially in the low field limit. The error bar of $T_{c}$ is determined by the half of difference between onset and offset. Since according to the {\color{blue}\cite{Fisher1962}}, d($\chi$$T$)/d$T$, d($\rho$)/dT behave like $C_p$ which gives transition temperature between onset and offset points, we use average of onset and offset value of d($\chi$$T$)/d$T$ and d($\rho$)/dT as $T^*$. For $T^{*}$, although the feature is much clearer for Cr substitution than it was for Mn, Ni or Co substitutions {\color{blue}\cite{Meier2018,Mingyu01}}, the features in \textit{M}(\textit{T}) and \textit{R}(\textit{T}) are still somewhat subtle in low substitution level. We infer $T^{*}$ as the average of onset and offset value and use the half of the difference between onset and offset as the error.

\begin{figure}
	\centering
	\includegraphics[width=2.6\columnwidth]{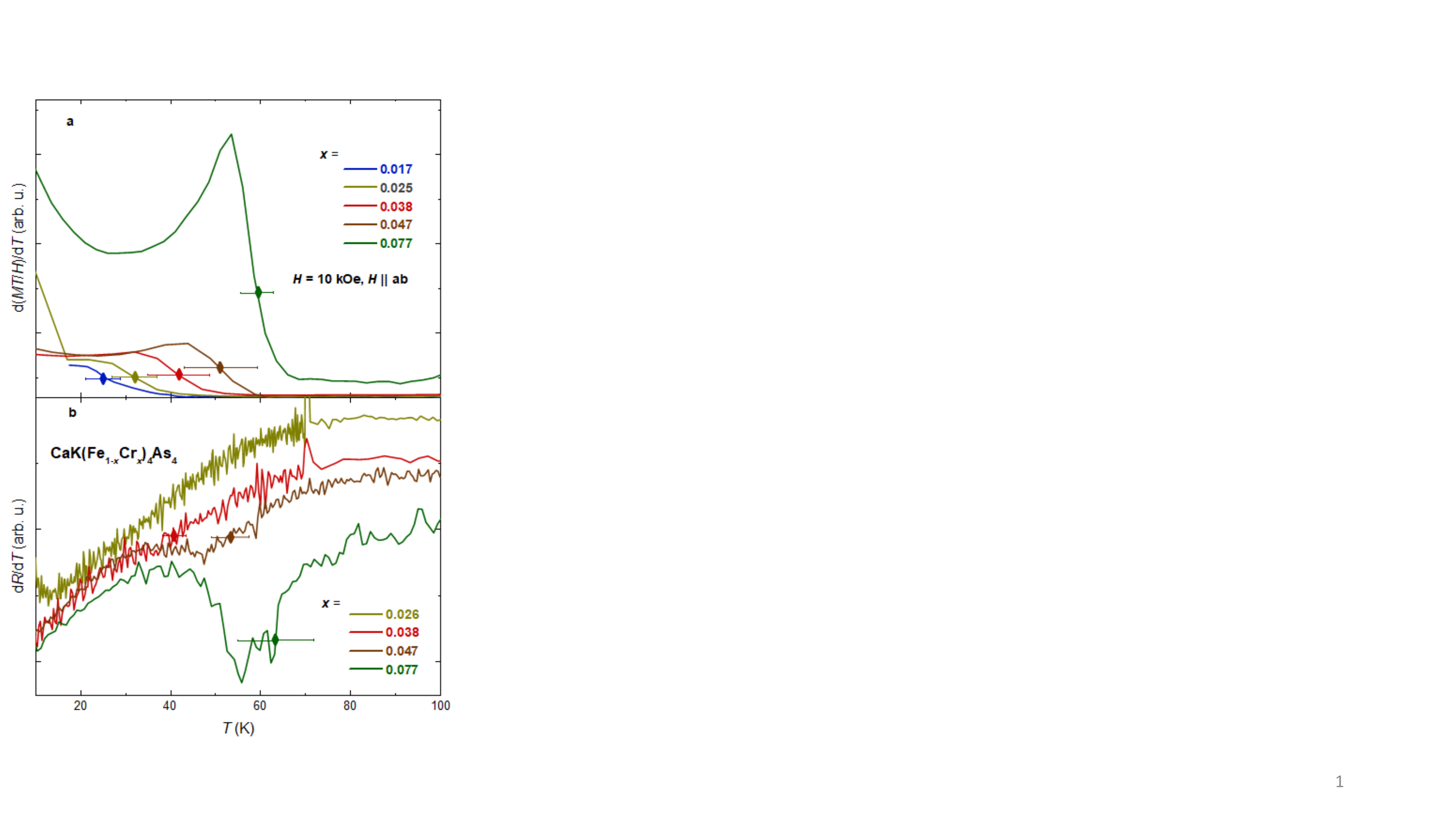}
	\caption{The $T^*$ anomaly appears clearly as a step in both plot d($MT/H$)/d$T$ and the derivative of resistance, d\textit{R}/d\textit{T}. Only the data above $T_c$ are plotted. Rhombuses symbols show the transition temperature of $T^*$ and error bars come from the criteria introduced above.\label{figure 14}}
\end{figure}
d\textit{R}(\textit{T})/d\textit{T} and d\textit{M}(T)/d\textit{T} data for several different \textit{x}-values are shown in figure \ref{figure 14}, showing good agreement between the position of the $T^{*}$ features.
\begin{figure}
	\includegraphics[width=1.5\columnwidth]{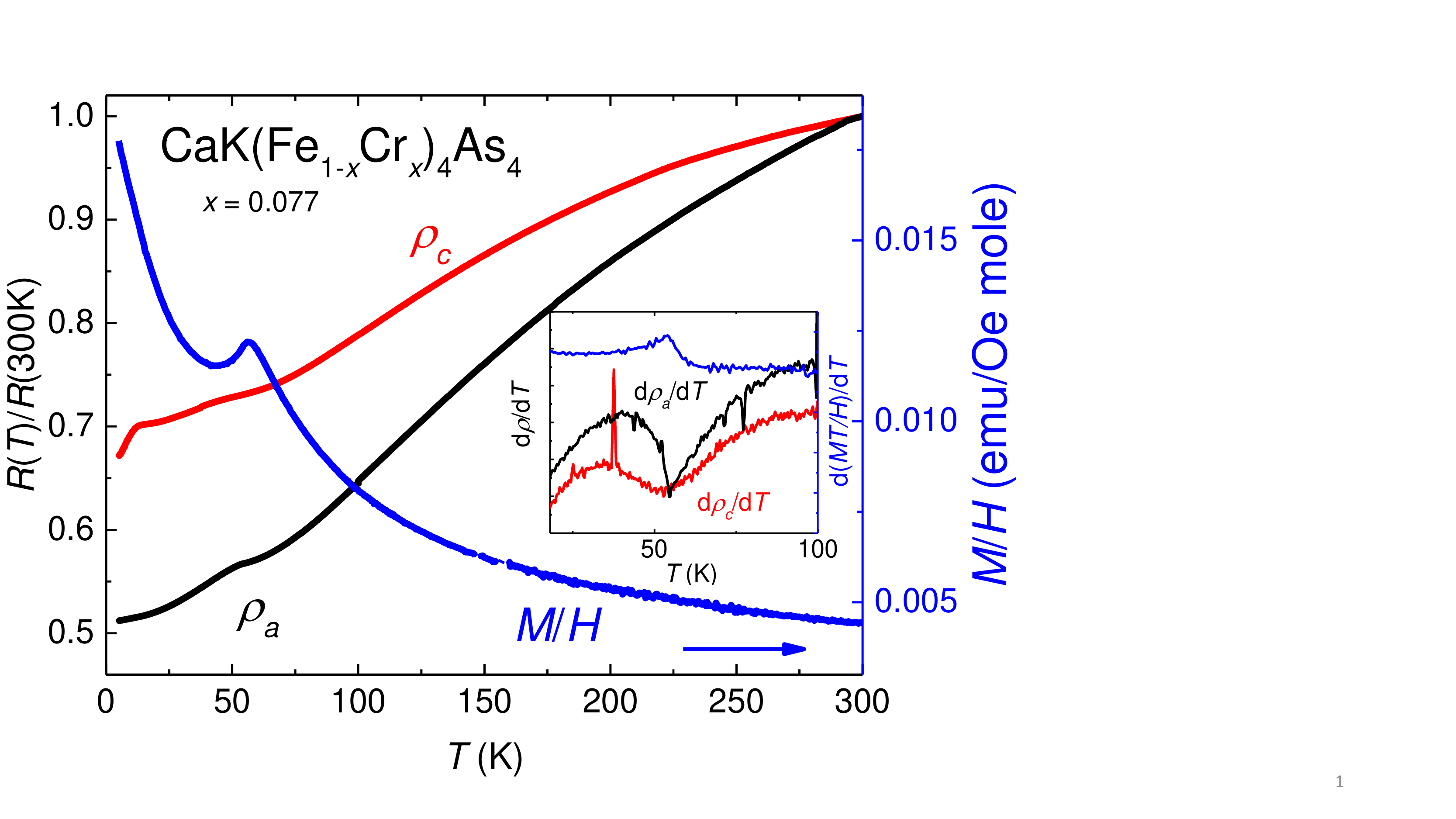}		
	\caption{Temperature dependence of normalized resistance, $\textit{R}(\textit{T})/\textit{R}$(300~K), of CaK(Fe$_{1-x}$Cr$_{x}$)$_{4}$As$_{4}$ single crystals with $x$ = 0.077 for electrical currents along $a$-axis (black) and along $c$-axis (red). Magnetic susceptibility shows clearly distinguished anomaly, similar to the features observed in raw resistivity data for both current directions. Inset shows resistivity derivatives plotted against d($MT$/$H$)/d$T$ (blue). The onset of the feature in the d($MT$/$H$)/d$T$ curve at $\sim$50~K is accompanied by the clear feature in both $c$- and $a$-axis resistivity derivative.
		\label{resfigrhoc3}}
\end{figure}
\begin{figure}[H]
	\includegraphics[width=1.5\columnwidth]{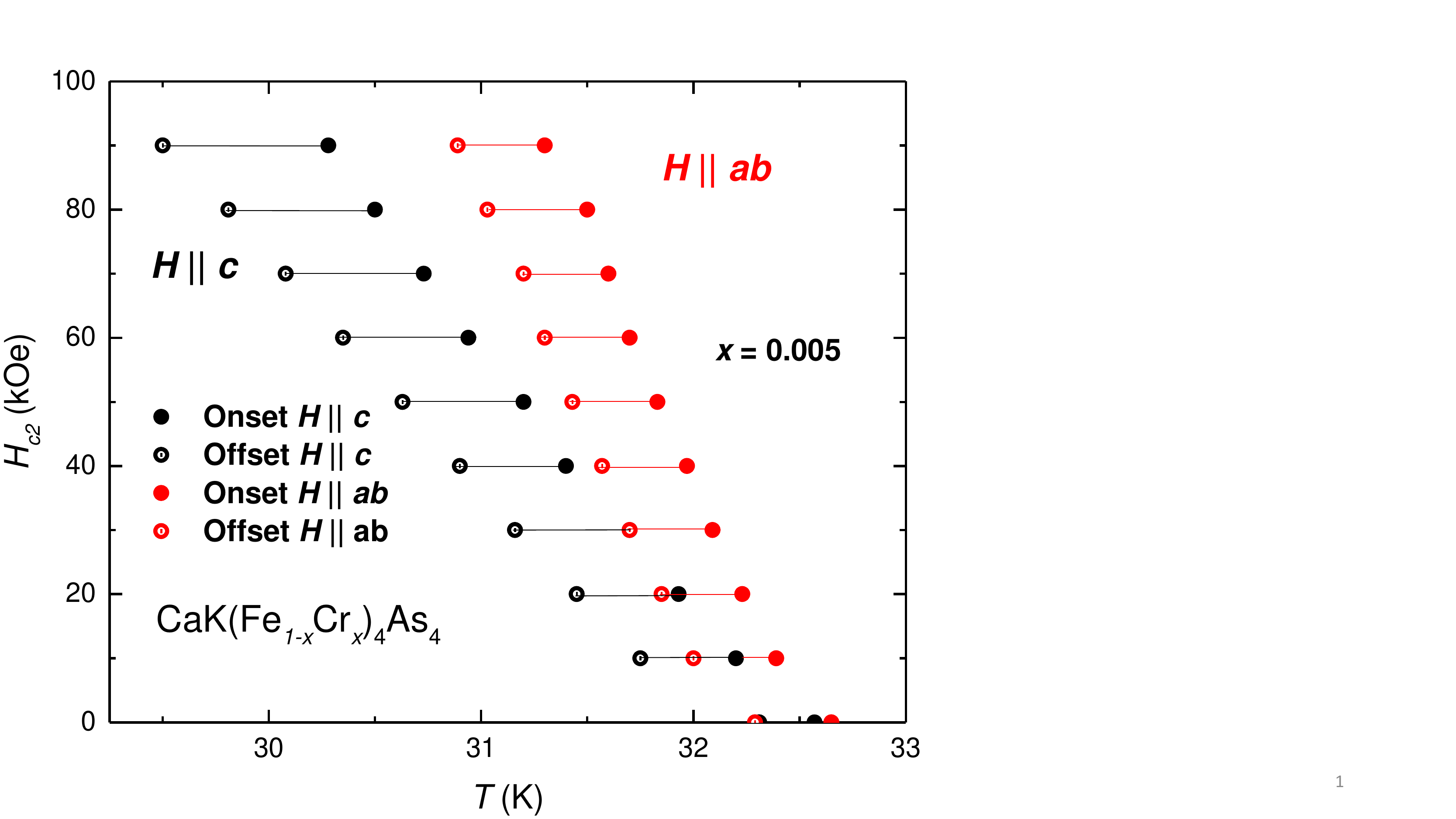}
	\caption{Anisotropic $H_{c2}(T)$ data determined for two single crystalline samples of $x$$_{EDS}$=0.005 CaK(Fe$_{1-x}$Cr$_{x}$)$_{4}$As$_{4}$ using onset criterion (solid) and offset criterion (hollow) inferred from the temperature-dependent electrical resistance data.   \label{figure81}}
\end{figure}

\begin{figure}[H]
	\includegraphics[width=1.5\columnwidth]{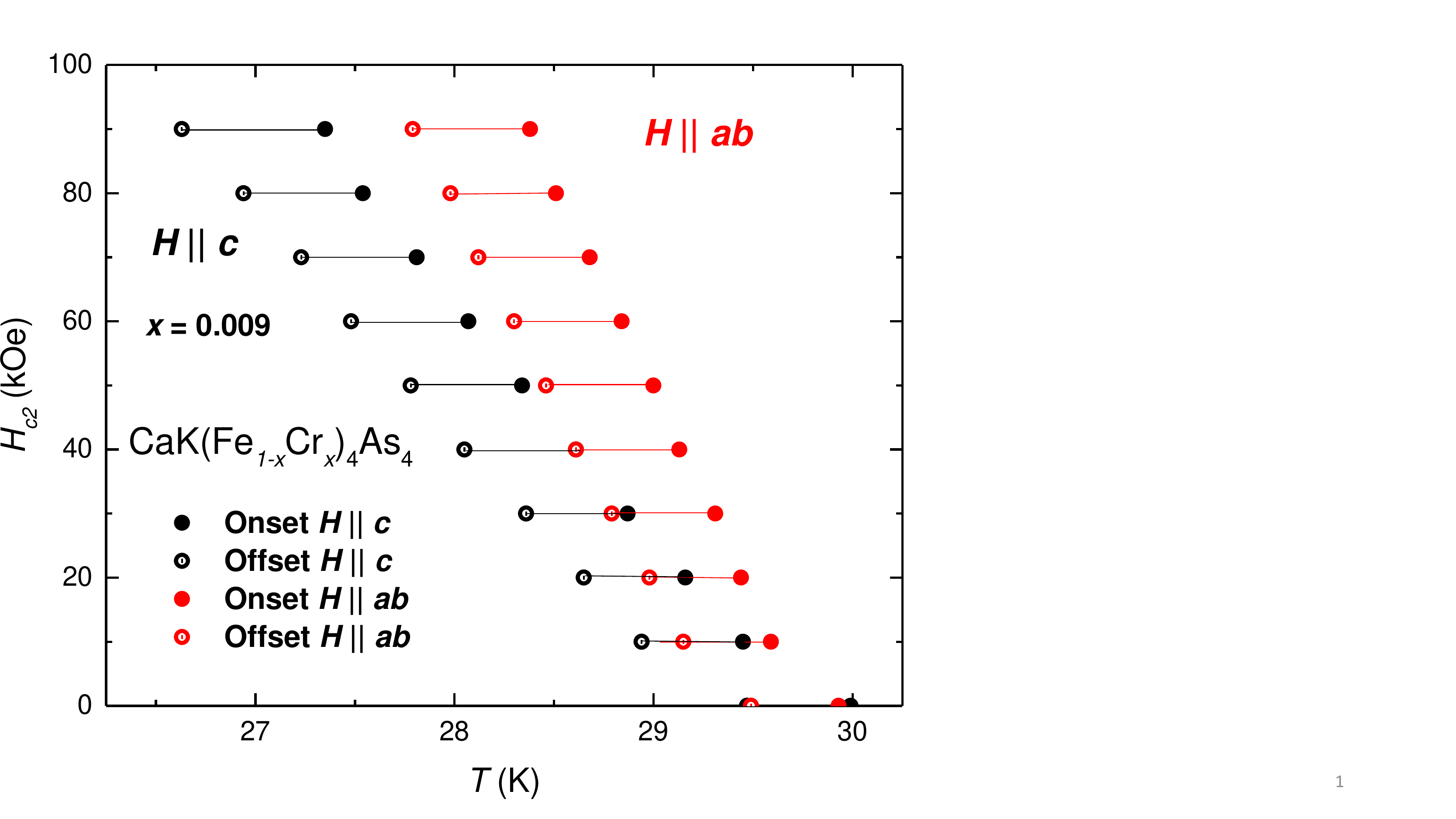}	
	\caption{Anisotropic $H_{c2}(T)$ data determined for two single crystalline samples of $x$$_{EDS}$=0.009 CaK(Fe$_{1-x}$Cr$_{x}$)$_{4}$As$_{4}$ using onset criterion (solid) and offset criterion (hollow) inferred from the temperature-dependent electrical resistance data.   \label{figure82}}
\end{figure}

\begin{figure}[H]
	\includegraphics[width=1.5\columnwidth]{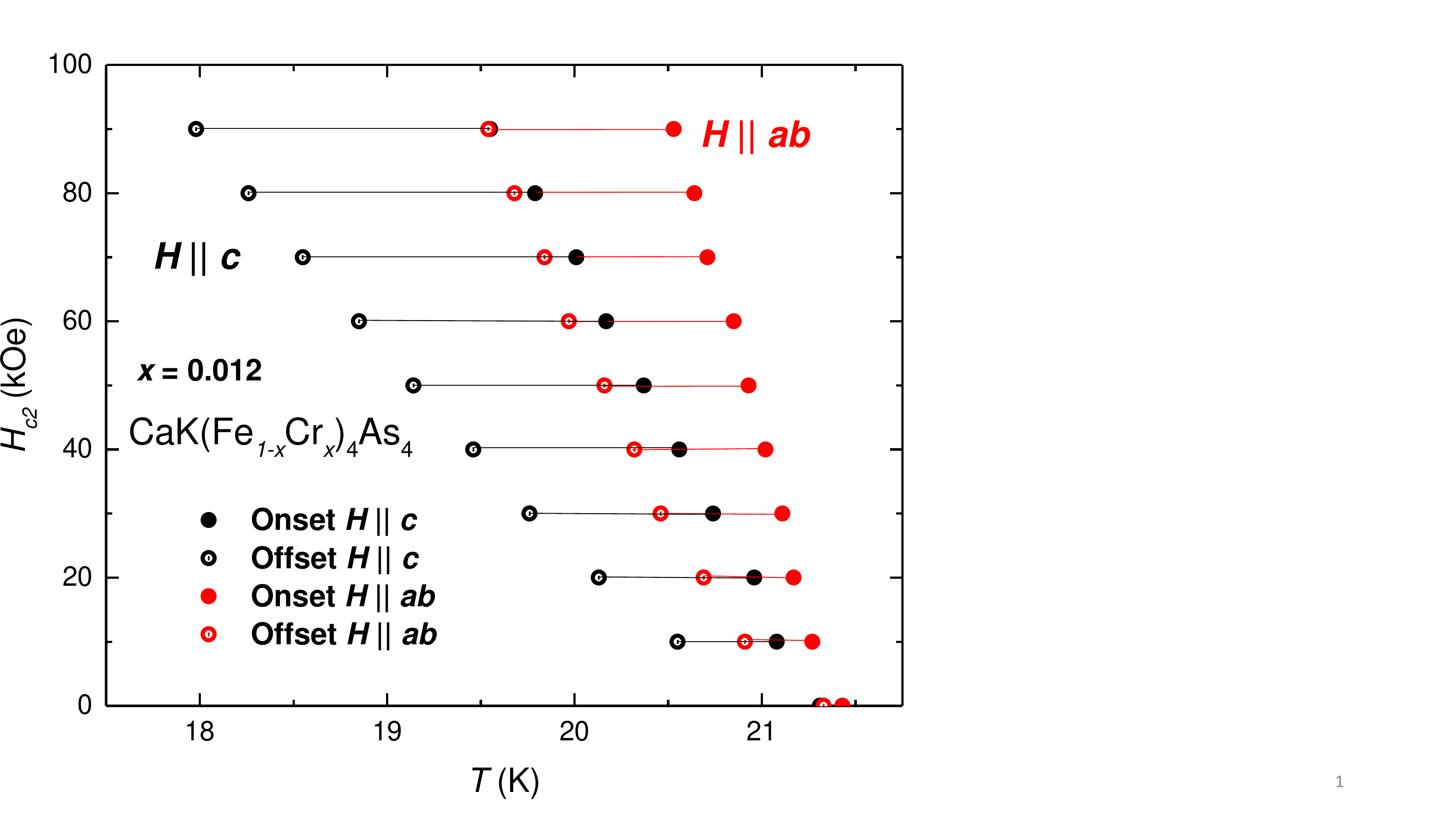}	
	\caption{Anisotropic $H_{c2}(T)$ data determined for two single crystalline samples of $x$$_{EDS}$=0.012 CaK(Fe$_{1-x}$Cr$_{x}$)$_{4}$As$_{4}$ using onset criterion (solid) and offset criterion (hollow) inferred from the temperature-dependent electrical resistance data.   \label{figure83}}
\end{figure}

\begin{figure}[H]
	\includegraphics[width=1.5\columnwidth]{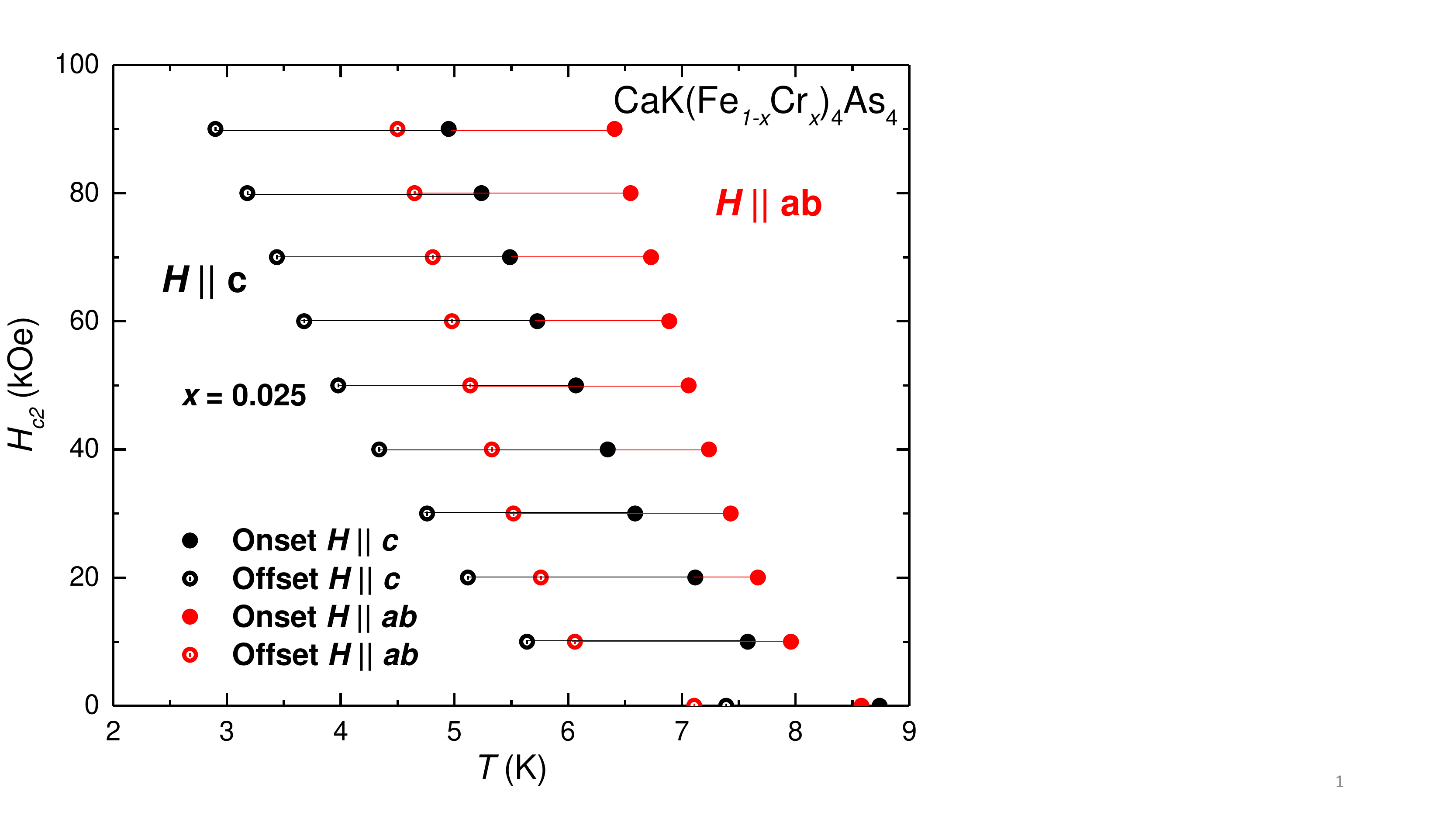}	
	\caption{Anisotropic $H_{c2}(T)$ data determined for two single crystalline samples of $x$$_{EDS}$=0.025 CaK(Fe$_{1-x}$Cr$_{x}$)$_{4}$As$_{4}$ using onset criterion (solid) and offset criterion (hollow) inferred from the temperature-dependent electrical resistance data.   \label{figure84}}
\end{figure}

Figures \ref{figure81} - \ref{figure84} present $H_{c2}(T)$ curves which is obtained from $R$($T$) data for fixed applied fields and the criteria shown in Figs \ref{figure7}a of CaK(Fe$_{1-x}$Cr$_{x}$)$_{4}$As$_{4}$ single crystals for $x$ = 0.005, 0.009, 0.012, 0.017 and 0.025.

\end{document}